\documentclass[preprint,12pt]{elsarticle}
\usepackage{etoolbox}
\makeatletter
\def\ps@pprintTitle{%
	\let\@oddhead\@empty
	\let\@evenhead\@empty
	\def\@oddfoot{\reset@font\hfil\thepage\hfil}
	\let\@evenfoot\@oddfoot
}
\makeatother
\usepackage{chngcntr}
\counterwithout{figure}{section}
\counterwithout{table}{section}
\usepackage{graphics}
\usepackage{amsmath, amsthm, amssymb}
\usepackage{natbib}
\usepackage[colorlinks=true]{hyperref}
\usepackage{lscape}
\usepackage[utf8]{inputenc} 
\usepackage{amsthm}

\usepackage{enumerate}
\usepackage{mathrsfs}
\usepackage{setspace}
\usepackage{multirow}
\usepackage{graphicx}
\usepackage{amsfonts}
\usepackage{verbatim}
\usepackage{amssymb}
\usepackage{amsthm}
\usepackage{caption}
\usepackage{subcaption}
\usepackage{multirow,bigstrut,bigdelim}

\theoremstyle{plain}

\theoremstyle{remark}

\numberwithin{equation}{section}
\usepackage[mathscr]{euscript}
\usepackage{geometry} 
\usepackage{graphicx,lscape} 
\usepackage{booktabs}
\usepackage{array}
\usepackage{paralist} 
\usepackage{verbatim}
\makeatletter
\renewcommand{\fnum@figure}{\textbf{Fig. \thefigure}}
\makeatother
\usepackage{caption}
\usepackage[labelsep=space]{caption}
\biboptions{comma,round,authoryear}

\begin{document}
	
	\begin{frontmatter}

        \title{A Bayesian promotion time cure model with current status data}
		\author{Pavithra Hariharan} 
		\author{P. G. Sankaran } 
		\address{Department of Statistics, Cochin University of Science and Technology, Cochin 682 022, Kerala, India\\ Corresponding author email: pavithrahariharan97@gmail.com }

				\begin{abstract}
   Analysis of lifetime data from epidemiological studies or destructive testing often involves current status censoring, wherein individuals are examined only once and their event status is recorded only at that specific time point. In practice, some of these individuals may never experience the event of interest, leading to current status data with a cured fraction. Cure models are used to estimate the proportion of non-susceptible individuals, the distribution of susceptible ones, and covariate effects. Motivated from a biological interpretation of cancer metastasis, promotion time cure model is a popular alternative to the mixture cure rate model for analysing such data. The current study is the first to put forth a Bayesian inference procedure for analysing current status data with a cure fraction, resorting to a promotion time cure model. An adaptive Metropolis-Hastings algorithm is utilised for posterior computation.  Simulation studies prove our approach's efficiency, while analyses of lung tumor and breast cancer data illustrate its practical utility. This approach has the potential to improve clinical cure rates through the incorporation of prior knowledge regarding the disease dynamics and therapeutic options.
			\end{abstract}
		
  
  \end{frontmatter}

\newpage
\section{Introduction}
Modelling of lifetime data becomes challenging due to the complexities arising from incomplete data, attributed to different forms of censoring. Let $T>0$ be a random variable quantifying the time taken for an event to take place. Current status censoring conceals $T$, but allows for observation of a single random monioring time $U>0$ and a random interval where $T$ falls.
Under a conventional current status data model, the data observable from each subject is $\left(U, \delta\right)$, where $\delta=I\left(T\leq U\right)$ that takes value 1 for events within $\left[0,U\right]$ and 0 for events within $\left(U,\infty\right]$. This category of data is identified as current status data (alternatively termed type I interval censored data), due to its nature of revealing only the present condition of the observed individual. These data might be preferred for destructive testing and epidemiological research, where repeated evaluations pose significant challenges due to practical, ethical, and logistical constraints. These constraints include limited resources, the invasive nature of testing, or the inherent design of the study, which makes continuous or frequent monitoring impractical
(\citeauthor{groeneboom1992information} \citeyear{groeneboom1992information}; \citeauthor{jewell2003current} \citeyear{jewell2003current}).

Examples comprise data on uterine fibroid development status at the ultrasound examination time arising out of a ``Right from the Start" prospective cohort study held at three different states of United States (\citeauthor{laughlin2009prevalence} \citeyear{laughlin2009prevalence}), data on renal recovery status of acute kidney injured at the time of discharge from University of Michigan Hospital (\citeauthor{heung2012fluid} \citeyear{heung2012fluid}; \citeauthor{al2022efficient} \citeyear{al2022efficient}), and gonorrhea status data from a Nebraska Public Health Laboratory survey (\citeauthor{li2021simulation} \citeyear{li2021simulation}).
There has been substantial interest in the analysis of current-status data,
see \cite{diamond1986proportional}, \cite{jewell1990statistical},  \cite{andersen1995nonparametric},
\cite{li2024regression}, \cite{wang2024martingale}, and \cite{wu2024deep}.
Some recent studies on Bayesian modelling utilising current status data belong to  \cite{cai2011bayesian}, \cite{das2024bayesian}, and \cite{paulon2024bayesian}.

The aforementioned studies postulate that each person in the study encounters the event eventually, given adequate follow-up time. However, in today's healthcare system, wherein many fatal diseases are recoverable, there may exist
certain individuals immune to the event, considered
cured. This can cause the typical survival curve to plateau rather than decline to zero.  Cure models are specifically designed for analysing such data by estimating cured proportions, understanding the lifetime distribution of uncured patients, and assessing the impact of covariates on lifetime (\citeauthor{maller1996survival} \citeyear{maller1996survival}).

There exist two classifications for cure models: the traditional two-component mixture cure model and the relatively recent promotion time cure model. The mixture cure model, pioneered by \cite{boag1949maximum}, further refined by \cite{berkson1952survival} and later explored by  
\cite{wang2020bayesian} and \cite{felizzi2021mixture} is a prevalent model for estimating the survival rate of untreated patients and cure rate of a treatment simultaneously. As a widely favoured substitute to the mixture cure model, \cite{tsodikov1996stochastic} and  \cite{tsodikov1998proportional} offered and investigated the promotion time cure model,
\begin{equation}\label{1.1}
    S_{pop}(t)=\exp\left[-\beta F(t)\right],
\end{equation}
where $\beta>0$ and $F(\cdot)$ is a proper baseline distribution function. It is also known as bounded cumulative hazard model since cumulative hazard $-\log S_{pop}(t)=\beta F(t)<\infty$. The model can be best understood through the following seminal biological interpretation
of \eqref{1.1}, based on a biological mechanism observed in cancer patients after treatment. The number of active carcinogenic cells remaining after treatment be denoted by $N$ and modelled by a Poisson distribution having mean $\beta$. 
$C_i$  denotes the time taken by the $i^{th}$ cell to develop an identifiable cancerous growth, with $T$  as their minimum. Assuming $C_i$s are independently distributed with common cumulative distribution function $F(t)$, $ S_{pop}(t)$ is the survival function of $T$, signifying the probability of being cancer-free by time $t$. 

\cite{chen1999new} modified \eqref{1.1} for modelling the population survival function of an individual pertaining to covariate $\textbf{X}=(1,X_1, \dots, X_k)'$,  as
\begin{equation}\label{1.2}
S_{pop}\left(t|\textbf{X}\right)=P\left(T>t|\textbf{X}\right)=
    \exp\left[-e^{\boldsymbol{\theta}'\textbf{X}}F(t)\right],
\end{equation}
by letting $\beta=e^{\boldsymbol{\theta}'\textbf{X}}$, where $\boldsymbol{\theta}=(\theta_{0},\theta_{1}, \dots, \theta_{k})'$ represents the vector of regression parameters. The immune or cured proportion according to model \eqref{1.2} equals
\begin{equation}\label{1.3}
    p(\textbf{X})=\lim_{t\rightarrow \infty}S_{pop}\left ( t|\textbf{X} \right )=\exp\left[-e^{\boldsymbol{\theta}'\textbf{X}}\right]>0.
\end{equation}
It may be noted that, under the model \eqref{1.2}, the covariates influence both the survival of patients who remain uncured and the probability of being cured. 
From a Bayesian perspective, the promotion time cure model outperforms the mixture cure model by yielding proper posterior distributions even with non-informative improper priors. Additionally, it features a proportional hazards structure and is biologically motivated, making it better suited for describing cancer relapse processes 
(\citeauthor{legrand2019cure} \citeyear{legrand2019cure}). Within the framework of the promotion time cure model, a nonparametric method for modelling the effects of covariates is introduced by \cite{chen2018promotion}  and  a semiparametric version of the model that accounts for nonlinear effects of a continuous covariate is explored by \cite{lin2024promotion}. Further, the efficiency and identifiability of the model have been examined and investigated by \cite{portier2017efficiency} and \cite{lambert2019estimation} respectively. For recent research on promotion time cure model, one could refer to  \cite{gomez2023general}, \cite{lin2024promotion}, and \cite{pal2023semiparametric} among many others. Some notable Bayesian inference procedures employing the promotion time cure model include those by \cite{rahimzadeh2016promotion}  and \cite{gressani2018fast}.

Limited literature exists on modelling current status data arising out of a population having an immune subgroup.  Notable frequentist contributions include mixture cure models by \cite{lam2005semiparametric} and \cite{liu2017regression}, a generalized linear model by \cite{ma2009cure}, an additive risk model by \cite{ma2011additive}, a non-mixture cure model by \cite{wang2020efficient}, and  transformation cure models by \cite{diao2019class} and \cite{lam2021marginal}. The Bayesian approach utilising Markov Chain Monte Carlo (MCMC) methods, offers advantages over frequentist approaches by incorporating prior information and providing exact inference without relying on asymptotic approximations. To date, the only Bayesian cure models developed for interval-censored data are those by \cite{ahmed2021bayesian} and \cite{pan2023bayesian}.
 However, a Bayesian promotion time cure model for modelling current status data with covariates is not yet developed. Inspired by this research gap, we have devised a Bayesian method intended to analyse current status data, relying on a promotion time cure model.

The paper adheres to the following structure. Section \ref{sec2} sets forth the Bayesian inference procedure, followed by techniques for model validation and comparison. Efficiency of the suggested method is inquired into by means of simulation studies and utility is exemplified with real datasets in Section \ref{sec3} and \ref{sec4} respectively. To conclude, certain observations are presented in Section \ref{sec5}. 


\section{Bayesian Inference Procedure}\label{sec2}

Consider the promotion time cure model \eqref{1.2}. Our objective is to estimate the distribution of uncured population through $F(t)$, covariate effects through $\boldsymbol{\theta}$ and the cured proportion $p(\textbf{X})$ given by \eqref{1.3}. Under current status censoring, assume non-informative censoring through the notion that, given $\textbf{X}$, $T$ and $U$ are independent. Further suppose that the monitoring time distribution depends on none of the parameters of interest.  Presuming $n$ subjects in the study who are independent,  we obtain the data, $\mathscr{D}$=
$\{(U_{i},\delta_{i},\textbf{X}_{i}):i=1, \dots, n\}$. With $\mathscr{D}$, the associated likelihood is established as
\begin{equation}\label{2.1}
    L(\boldsymbol{\theta}, F(\cdot)|\mathscr{D})=\prod_{i=1}^{n}S_{pop}\left ( U_i|\textbf{X}_i\right )^{1-\delta_i}\left ( 1-S_{pop}\left ( U_i|\textbf{X}_i\right ) \right )^{\delta_i}.
\end{equation}
Given assumption \eqref{1.2}, \eqref{2.1} transforms into
\begin{equation}\label{2.2}
     L(\boldsymbol{\theta}, F(\cdot)|\mathscr{D})=\prod_{i=1}^{n}\exp\left[-e^{\boldsymbol{\theta}'\textbf{X}_i}F(U_i)(1-\delta_i)\right]\left ( 1-\exp\left[-e^{\boldsymbol{\theta}'\textbf{X}_i}F(U_i)\right] \right)^{\delta_i}.
\end{equation}

\subsection{Prior Distributions}\label{ss2.1}
Bayesian analysis builds on prior knowledge about the parameters in a statistical model, represented by the prior distributions, even before considering any data.

One may observe from \eqref{2.2} that merely the values of the continuous distribution function $F(\cdot)$ at $U_{i}$'s affect the likelihood function. Therefore, without loss of generality, one can focus only on the estimation of $F(\cdot)$ within the class of all non-decreasing step functions having discontinuities at distinct monitoring times, say $0=s_{0}<s_{1}<s_{2}<\dots<s_{n_{0}}$. As presented by \cite{sun2006statistical}, consider the step functions of the form,
\begin{equation}\label{2.3}
    F_{s}(t)=1-\prod_{l:s_{l}\leq t}
    \exp\left(-e^{\eta_{l}}\right),
\end{equation}
with $\boldsymbol{\eta} =(\eta_{1}, \dots,\eta_{n_0})'$, where $\eta_{l}=\log(-\log(r_{l}))$ and $r_{l}=\frac{1-F_{s}(s_{l})}{1-F_{s}(s_{l-1})}$, for $l=1,\dots,n_{0}$. For prior of 
 $\boldsymbol{\eta}$, an $n_{0}$-variate normal distribution with mean $\boldsymbol{\mu}=(\mu_{1},\dots,\mu_{n_{0}})'$ is chosen. As its variance covariance matrix, $\boldsymbol{\Sigma_{\boldsymbol{\eta}}}$,  an $n_{0}\times n_{0}$ matrix with non-zero
non-diagonal elements is selected, to account for the dependence of $\eta_l$;\quad$l=1,\dots,n_{0}$ with its adjacent components. Thus, $\boldsymbol{\eta} \sim N_{n_{0}}(\boldsymbol{\mu},\boldsymbol{\Sigma_{\boldsymbol{\eta}}})$ with the probability density function:
\begin{equation*}
 \pi_{\boldsymbol{\eta}}(\boldsymbol{\eta})=\frac{1}{(2\pi)^{n_{0}/2}}|\boldsymbol{\Sigma_{\boldsymbol{\eta}}}|^{-1/2}e^{\frac{-1}{2}(\boldsymbol{\eta}-\boldsymbol{\mu})'\boldsymbol{\Sigma_{\boldsymbol{\eta}}}^{-1}(\boldsymbol{\eta}-\boldsymbol{\mu})},
\end{equation*}
where $\boldsymbol{\mu}$ and $\boldsymbol{\Sigma_{\boldsymbol{\eta}}}$ can be either known or estimated by the experimenter based on their expertise.

Assuming the independence of regression coefficients with real support, a $(k+1)$-variate normal distribution is adopted as the prior for $\boldsymbol{\theta}$.  This distribution has mean vector $\boldsymbol{\tau}=(\tau_{0}, \tau_{1},\dots,\tau_{k})'$ and a $(k+1) \times (k+1)$ variance-covariance matrix $\boldsymbol{\Sigma_{\boldsymbol{\theta}}}$, having zeroes as non-diagonal elements, signifying independence between the coefficients. Therefore, $\boldsymbol{\theta} \sim N_{k+1}(\boldsymbol{\tau},\boldsymbol{\Sigma}_{\boldsymbol{\theta}})$ characterised by its probability density function
\begin{equation*}
 \pi_{\boldsymbol{\theta}}(\boldsymbol{\theta})=\frac{1}{(2\pi)^{(k+1)/2}}|\boldsymbol{\Sigma_{\boldsymbol{\theta}}}|^{-1/2}e^{\frac{-1}{2}(\boldsymbol{\theta}-\boldsymbol{\tau})'\boldsymbol{\Sigma_{\boldsymbol{\theta}}}^{-1}(\boldsymbol{\theta}-\boldsymbol{\tau})}, 
\end{equation*}
where $\boldsymbol{\tau}$ and $\boldsymbol{\Sigma_{\boldsymbol{\theta}}}$ are chosen by the experimenter to incorporate prior knowledge or uncertainty regarding the regression parameters. 

Additionally, the independence between $\boldsymbol{\eta}$ and $\boldsymbol{\theta}$ is  assumed, given the covariate vector $\textbf{X}$.

\textbf{Remark 2.1}\
The priors are selected based on experimental knowledge, with options ranging from non-informative priors that offer inferential priority to data, to informative priors when prior knowledge is available.

\subsection{Posterior Computation}
The likelihood is coupled with the priors to form the posterior distribution, encapsulating improved beliefs about the parameters after observing the data.

Employing \eqref{2.3}, the likelihood function \eqref{2.2}
is rewritten in terms of $\boldsymbol{\theta}$ and $\boldsymbol{\eta}$ by
\begin{equation}\label{3.4}
   \begin{aligned}
      L( \boldsymbol{\theta},\boldsymbol{\eta}|\mathscr{D})
      &=\prod_{i=1}^{n}\exp\left[-e^{\boldsymbol{\theta}'\textbf{X}_i}\left(1-\prod_{l:s_{l}\leq U_i}
    \exp\left(-e^{\eta_{l}}\right)\right)(1-\delta_i)\right]\\
    &\times \left ( 1-\exp\left[-e^{\boldsymbol{\theta}'\textbf{X}_i}\left(1-\prod_{l:s_{l}\leq U_i}
    \exp\left(-e^{\eta_{l}}\right)\right)\right] \right )^{\delta_i}.
\end{aligned}
\end{equation}
 
Taking into account the dataset $\mathscr{D}$ and postulates specified in Subsection \ref{ss2.1}, concerning the prior distributions of $\boldsymbol{\theta}$ and $ \boldsymbol{\eta}$ as well as their independence, posterior distribution $\pi ^{*}(\cdot)$ is expressed as
\begin{equation}\label{2.5}
          \pi ^{*}(\boldsymbol{\theta}, \boldsymbol{\eta}|\mathscr{D})
          \propto \pi_{\boldsymbol{\theta}}(\boldsymbol{\theta})\pi_{\boldsymbol{\eta}}(\boldsymbol{\eta})L(\boldsymbol{\theta}, \boldsymbol{\eta}|\mathscr{D}).
\end{equation}
Adopting the squared error loss function, marginal posterior means of the parameters provide the Bayes estimators as  
\begin{equation}\label{A.2}
\Tilde{\theta}_{\nu }
= E_{\pi_{\nu }^{*}}(\theta_{\nu }|\mathscr{D})=\int_{\theta_{0}}\int_{\theta_{1}}\dots \int_{\theta_{\nu }}\dots\int_{\theta_{k}}\theta_{\nu }\pi_{\boldsymbol{\theta}}^{*}(\boldsymbol{\theta}|\mathscr{D})d\theta_{0}\,d\theta_{1}\dots d\theta_{\nu }\dots d\theta_{k};\nu=0,1,\dots,k,
\end{equation}
and
\begin{equation}\label{A.4}
 \begin{aligned}
\Tilde{\eta}_{l}
&= E_{\pi_{l}^{*}}(\eta_{l}|\mathscr{D})=\int_{\eta_{1}}\int_{\eta_{2}}\dots \int_{\eta_{l }}\dots\int_{\eta_{n_0}}\eta_{l}\pi_{\boldsymbol{\eta}}^{*}(\boldsymbol{\eta}|\mathscr{D})d\eta_{1}\,d\eta_{2}\dots d\eta_{l}\dots d\eta_{n_0};l=1,\dots,n_0,
\end{aligned}   
\end{equation}
where $\pi_{\boldsymbol{\theta}}^{*}(\boldsymbol{\theta}|\mathscr{D})$ and $ \pi_{\boldsymbol{\eta}}^{*}(\boldsymbol{\eta}|\mathscr{D})$ denote the marginal posterior densities of $\boldsymbol{\theta}$ and $\boldsymbol{\eta}$ respectively.
 
An approach for estimating $F(t)$ 
utilising the Bayes estimators $\Tilde{\eta}_{l}$, $l=1,\dots,n_{0}$, in 
\eqref{A.4}, is proposed through the formulation;
\begin{equation}\label{2.6}
    \Tilde F_{s}(t)=1-\prod_{l:s_{l}\leq t}
    \exp\left(-e^{\Tilde{\eta}_{l}}\right).
\end{equation}
The estimator for the population survival function of a subject given the vector of covariates $\textbf{X}$ is suggested as
\begin{equation}\label{2.7}
    \Tilde{S}_{pop}\left(t|\textbf{X}\right)=
    \exp\left(-e^{\Tilde{\boldsymbol{\theta}}'\textbf{X}}\left [ 1-\prod_{l:s_{l}\leq t}
    \exp\left(-e^{\Tilde{\eta}_{l}}\right) \right ]\right).
\end{equation}
With the vector of covariates $\textbf{X}$ and Bayes estimators from \eqref{A.2}, $\boldsymbol{\Tilde{\theta}}=(\Tilde{\theta}_{0},\Tilde{\theta}_{1},\dots,\Tilde{\theta}_{k})'$ in hand, the proposed estimator for the cure fraction is expressed as
\begin{equation}\label{2.8}
     \Tilde{p}(\textbf{X})=\lim_{t\rightarrow \infty}\Tilde{S}_{pop}\left ( t|\textbf{X} \right )=\exp\left(-e^{\Tilde{\boldsymbol{\theta}}'\textbf{X}}\right).
\end{equation} 
Theoretical assessment of \eqref{A.2} and \eqref{A.4} is complicated, prompting the utilisation of MCMC methods.

\subsection{Posterior Simulation}\label{ss2.3}
The conditional posterior densities of parameters $\theta_{\nu};\nu=0,1,\dots,k$ and $\eta_{l};l=1,\dots,n_0$ do not possess closed forms and 
 Gibbs sampling ceases to be applicable. Therefore,  
an adaptive Metropolis-Hastings (MH) algorithm (\citeauthor{haario1999adaptive} \citeyear{haario1999adaptive}) is deployed for posterior simulation utilising \textit{MHadaptive} package in \textit{R}-software with a few modifications. This sophisticated form of MH algorithm uses Gaussian proposal distribution with dynamic variance-covariance structure determined by process history. For more details and variations of the algorithm, see \cite{chauveau2002improving}, \cite{cai2008metropolis}, \cite{griffin2013adaptive}, and \cite{marnissi2020majorize}. 

The algorithm outlined in the subsequent steps generates a Markov chain
$(\boldsymbol{\theta}^{(m)},\boldsymbol{\eta}^{(m)})$
having $\pi^{*}(\cdot)$ as the approximate stationary distribution, where $\boldsymbol{\theta^{(m)}}=(\theta_{0}^{(m)},\theta_{1}^{(m)},\dots,\theta_{k}^{(m)})$ and $\boldsymbol{\eta}^{(m)}=(\eta_{1}^{(m)},\dots,\eta_{n_0}^{(m)})$. \\\\
(i) Formulate function \eqref{2.5} using the priors for $\boldsymbol{\theta}$, $\boldsymbol{\eta}$, and the dataset $\mathscr{D}$.\\
(ii) Set initial parameter values $(\boldsymbol{\theta}^{(0)},\boldsymbol{\eta}^{(0)})$  and calculate the parameter values $(\boldsymbol{\theta}^{(1)},\boldsymbol{\eta}^{(1)})$ that
maximise \eqref{2.5}, then set $m=1$.\\
(iii) Select the Gaussian proposal distribution, with the inverse of the observed Fisher information matrix evaluated at $(\boldsymbol{\theta}^{(1)},\boldsymbol{\eta}^{(1)})$ as the variance-covariance matrix and $(\boldsymbol{\theta}^{(m)'},$ $\boldsymbol{\eta}^{(m)'})'$ as mean.\\
(iv) Generate new values of parameters $\boldsymbol{\theta}^{(m)p}$ and $\boldsymbol{\eta}^{(m)p}$ out of the proposal distribution considered. Then choose $\omega$ randomly from $U(0,1)$.\\
(v) Calculate the transition probability $\phi ((\boldsymbol{\theta}^{(m)},\boldsymbol{\eta}^{(m)}),(\boldsymbol{\theta}^{(m)p},\boldsymbol{\eta}^{(m)p}))$ as the minimum of $\left \{1, \frac{\pi^{*}(\boldsymbol{\theta}^{(m)p},\boldsymbol{\eta}^{(m)p}|\mathscr{D})}{\pi^{*}(\boldsymbol{\theta}^{(m)},\boldsymbol{\eta}^{(m)}|\mathscr{D})} \right \}$. If $\log\,\,\omega$ is smaller than or equal to $ \log \,\,\phi((\boldsymbol{\theta}^{(m)},\boldsymbol{\eta}^{(m)}),(\boldsymbol{\theta}^{(m)p},\boldsymbol{\eta}^{(m)p}))$, update $\boldsymbol{\theta}^{(m+1)}=\boldsymbol{\theta}^{(m)p}$ and $\boldsymbol{\eta}^{(m+1)}=\boldsymbol{\eta}^{(m)p}$. Otherwise, set $\boldsymbol{\theta}^{(m+1)}=\boldsymbol{\theta}^{(m)}$ and  $\boldsymbol{\eta}^{(m+1)}=\boldsymbol{\eta}^{(m)}$.\\
(vi) Increase $m$ by one and execute steps (iii)-(v) for a predefined number of iterations, determined based on the Markov chain diagnostics. At specified intervals, the variance-covariance matrix of proposal distribution is updated adaptively using a portion of the previously generated values (\citeauthor{haario1999adaptive} \citeyear{haario1999adaptive}).\\
(vii) The values of parameters, followed by an adequate burn-in and suitable thinning, produce the posterior sample $(\boldsymbol{\theta}^{(m)},\boldsymbol{\eta}^{(m)})$;
$m=1,\dots,m_0$, that constitutes a nearly independent sample from a stationary distribution  approximating $\pi^{*}(\cdot)$.\\
(viii) Calculate \eqref{A.2} and \eqref{A.4} as 
\begin{equation}\label{2.9}
 \Tilde{\theta}_{\nu}=\frac{1}{m_0}\sum_{m=1}^{m_0}\theta_{\nu}^{(m)}, \nu=0,1,\dots, k,    
\end{equation}
where $\theta_{\nu}^{(m)}$ represents the $\nu^{th}$ element of the vector $\boldsymbol{\theta}^{(m)}$ for $m=1,\dots, m_0$. 
\begin{equation}\label{2.10}
   \Tilde{\eta}_{l}=\frac{1}{m_0}\sum_{m=1}^{m_0}\eta_{l}^{(m)}, l=1,\dots, n_0,
\end{equation}
where $\eta_{l}^{(m)}$ represents the $l^{th}$ element of the vector $\boldsymbol{\eta}^{(m)}$ for $m=1,\dots, m_0$.

The Ergodic theorem (\citeauthor{robert1999monte} \citeyear{robert1999monte}) ensures that the empirical averages \eqref{2.9} and  \eqref{2.10} converge to integrals \eqref{A.2} and  \eqref{A.4} respectively.

\subsection{Model Comparison and Validation}
Two common measures chosen for comparing multiple models from a Bayesian standpoint are the Log pseudo-marginal likelihood (LPML) and the Deviance Information Criterion (DIC) due to \cite{geisser1979predictive} and \cite{spiegelhalter2002bayesian} respectively. Various researchers have devised expressions for these, across different semiparametric regression models, utilising current status data (\citeauthor{wang2011semiparametric} \citeyear{wang2011semiparametric}; \citeauthor{hariharan2023bayesian} \citeyear{hariharan2023bayesian}; \citeauthor{hariharan2024semiparametric} \citeyear{hariharan2024semiparametric}). 

LPML is viewed as an indicator of the overall predictive performance of the data, with higher values suggesting better performance. It is computed using Bayesian cross validated residual or 
 conditional predictive ordinate values (CPO). $CPO_{i};i=1,\dots,n$ represents the predictive probability of $i^{th}$ observation taking into account leftover data; 
$\mathscr{D}^{(-i)}=\{(U_{j},\delta_{j},\textbf{X}_{j}): j=1,\dots, n, j\neq i\}$   assuming that the current model is valid. $CPO_{i}$ is  determied using
 \begin{equation*}
\begin{aligned}
CPO_{i}&=P(T_{i}\in (L_{i},R_{i}]|\mathscr{D}^{(-i)})\\
&=\left (E_{\pi^{*}}\left[\frac{1}{P(T_{i}\in (L_{i},R_{i}]|\boldsymbol{\theta},\boldsymbol{\eta})} \right] \right)^{-1},
\end{aligned}
\end{equation*}
$CPO_{i}$ due to \cite{chen2012monte} upon obtaining the MCMC sample $(\boldsymbol{\theta}^{(m)},\boldsymbol{\eta}^{(m)});m=1,\dots, m_{0}$ is
\begin{equation}
\begin{aligned}\label{2.11}
    CPO_{i}&=\left(\frac{1}{m_{0}}\sum_{m=1}^{m_{0}}\left[\frac{1}{P(T_{i}\in (L_{i},R_{i}]|\boldsymbol{\theta}^{(m)},\boldsymbol{\eta}^{(m)})}\right]\right)^{-1}\\
&=\left(\frac{1}{m_{0}}\sum_{m=1}^{m_{0}}\left[\frac{1}{\delta _{i}+(-1)^{\delta _{i}} \exp\left(-e^{\Tilde{\boldsymbol{\theta}^{(m)}}'\textbf{X}_i}\left [ 1-\prod_{l:s_{l}\leq U_i}
    \exp\left(-e^{\Tilde{\eta_{l}}^{(m)}}\right) \right ]\right)}\right]\right)^{-1}.
\end{aligned}
\end{equation}
Subsequently, LPML is estimated as
\begin{equation}\label{2.12}
   LPML=\sum_{i=1}^{n}\log\,\,CPO_i.
\end{equation}

DIC serves as a technique for comparing Bayesian models. It strikes a balance between model fit and complexity, with lower values suggesting a better fit while minimising complexity. The DIC is computed as the sum of the expected value of deviance $D(\boldsymbol{\theta}, \boldsymbol{\eta})$ and a penalty term $p_{D}$ for complexity, as  
\begin{equation*}
    DIC=E_{\pi^{*}}[D(\boldsymbol{\theta}, \boldsymbol{\eta})]+p_{D},
\end{equation*}
 where, \(E_{\pi^{*}}(\cdot)\) refers to the expected value with regard to  \(\pi^{*}(\cdot)\) specified in \eqref{2.5}, $D(\boldsymbol{\theta}, \boldsymbol{\eta})= -2 \log L(\boldsymbol{\theta}, \boldsymbol{\eta}|data) + c$, with $c$ a constant, and $p_{D}=E_{\pi^{*}}[D(\boldsymbol{\theta}, \boldsymbol{\eta})]-D[E_{\pi^{*}}(\boldsymbol{\theta}, \boldsymbol{\eta})]$. Therefore,
\begin{equation*}
    DIC=2E_{\pi^{*}}[D(\boldsymbol{\theta}, \boldsymbol{\eta})]-D[E_{\pi^{*}}(\boldsymbol{\theta}, \boldsymbol{\eta})],
\end{equation*}
which can be approximated using sample $(\boldsymbol{\theta}^{(m)},\boldsymbol{\eta}^{(m)});m=1,\dots, m_0$, as
\begin{equation}\label{2.13}
     DIC=2\Tilde{D}(\boldsymbol{\theta}, \boldsymbol{\eta})-D(\Tilde{\boldsymbol{\theta}},\Tilde{\boldsymbol{\eta}}).
\end{equation}
Here, $\Tilde{D}(\boldsymbol{\theta}, \boldsymbol{\eta})$
is the average deviance over $m_{0}$ samples and $\Tilde{\boldsymbol{\theta}}=(\Tilde{\theta}_{0},\Tilde{\theta}_{1},...,\Tilde{\theta}_{k})'$, $\Tilde{\boldsymbol{\eta}}=(\Tilde{\eta}_{1},...,\Tilde{\eta}_{n_{0}})$  are obtained according to equations \eqref{2.9} and \eqref{2.10} respectively.

CPO also serves as a valuable tool for assessing model adequacy. By plotting CPO values or scaled CPO values (normalised to the maximum CPO value) against observation indices or monitoring times $U_i;i=1,\dots,n$, analysts can evaluate model fit. Larger CPO values suggest better agreement between the model and observations, while low values (scaled CPO $<$ 0.01) imply outliers and poor model fit to the data (\citeauthor{congdon2005bayesian} \citeyear{congdon2005bayesian}). A model is believed satisfactory if its CPOs or scaled CPOs exhibit a random distribution without outliers, as corroborated by \cite{aslanidou1998bayesian} and \cite{sinha2004bayesian}.

\section{Simulation Studies}\label{sec3}
Evaluation of the proposed Bayesian estimation procedure in estimating model parameters is undertaken through simulation studies. Two covariates, $X_1$ distributed as a Bernoulli distribution having probability of success 0.5 and $X_2$ following a standard normal distribution are presumed to impact the event times. Five distinct parameter combinations representing both positive and negative covariate effects are assumed, and 500 datasets of identical size are generated for each combination. The Bayesian estimation procedure outlined is then applied to each dataset, yielding parameter estimates. Estimations are performed with sample sizes ($n$) of 200 and 500 and posterior summaries are provided subsequently. Consider  a  simple model
\begin{equation}\label{3.1}
S_{pop}\left(t|\textbf{X}\right)=
    \exp\left(-e^{\theta_0+\theta_1X_1+\theta_2X_2}\left(1-e^{-\frac{b}{a}(e^{at}-1)}\right)\right), 
\end{equation}
where $a,b>0$, $\theta_0,\theta_1,\theta_2$ are real numbers and $\textbf{X}=(1,X_1,X_2)'$.
A Gompertz distribution is used to model $F(t)$ with survivor function $S(t)=e^{-\frac{b}{a}(e^{at}-1)}$ 
 and \eqref{3.1} incorporates a proportion of immune subjects. 

Implementing the method discussed in \cite{oulhaj2014generating} for generating data from any improper distribution, current status data with a cured fraction is produced using the model \eqref{3.1}. 
To generate the required data, presume $a=0.5$, $b=1.1$ and keep $\boldsymbol{\theta}=(\theta_0,\theta_1,\theta_2)'$ fixed. For $i=1,\dots,n$, generate $\textbf{X}_i=(1,
X_{1i},X_{2i})'$, where $X_{1i}\sim B(1,0.5)$ and $X_{2i}\sim N(0,1)$. Compute $1-p(\textbf{X}_i)=1-\exp\left(-e^{\boldsymbol{\theta}'\textbf{X}_i}\right)$ and generate a random number $\lambda_i$ from $B(1,1-p(\textbf{X}_i))$. If $\lambda_i=0$, set $T_i=\infty$; if $\lambda_i=1$, generate $\chi_i$ from $Uniform(0,1)$ and compute the event time $T_i$ using,
    \begin{equation*}
        T_i=S^{-1}\left [ 1+\frac{1}{\exp(\boldsymbol{\theta}'\textbf{X}_i)} \log\left [ 1-\chi_i\left ( 1-\exp\left ( -\exp(\boldsymbol{\theta}'\textbf{X}_i) \right ) \right ) \right ]\right ].
    \end{equation*}

In scenario (1), a fixed censoring scheme is employed, selecting ten equidistant monitoring times 
$(s_{1},\dots,$ $ s_{10})=(0.3, 0.6, 0.9, 1.2, 1.5, 1.8, 2.1, 2.4, $ $2.7, 3.0)$, for ease of implementation. A multinomial random vector $(n_{1},\dots, n_{10})$ with all probabilities equal is generated, subject to $\sum_{l=1}^{10}n_{l}=n$. Decide the value of $U_i$ using the pattern: set $U_{i}$ equal to $s_{1}$ when $i$ is between 1 and $n_1$, equal to $s_2$ when $i$ is between $n_1 + 1$ and $n_1 + n_2$, and so on, until $U_i$ equals $s_{10}$ when $i$ is between $n_1 + n_2 + ... + n_9 + 1$ and $n$. For $i=1,\dots,n$, $\delta_{i}$ is set to 1 if $T_{i} $  is smaller than or equal to $U_{i}$ and otherwise set to zero. This method produces the current status data from a population with a cured fraction, represented as $\left\{(U_{i},\delta_{i},\textbf{X}_{i});i=1,\dots, n\right\}$.  Additionally, under scenario (2) involving a random censoring scheme, ten non-equidistant monitoring times are generated from $Uniform(0, 3)$, which is a more common occurrence in practice, 
and the entire process is then repeated.

To give inferential priority to data, vague prior  $N(1,10^{2})$ is chosen for $\theta_{0}$, $\theta_{1}$, and $\theta_{2}$. Using $\mu_{l}=\log\left[-\log\left(\frac{S(s_{l})}{S(s_{l-1})}\right)\right]$ for $l=1,\dots,10$, $\boldsymbol{\mu}=(-1.03,-0.88,-0.73,-0.58,$ $-0.43,-0.28,-0.13,0.02,0.17,0.32)'$ is obtained. Noting that 
$\eta_l;l=1,\dots,n_0$, by definition, has the highest dependence with adjacent components and the lowest with distant ones, an appropriate choice for $\boldsymbol{\Sigma}_{\eta}$ is a first-order autoregressive structure $\boldsymbol{\Sigma}_{n_0}(\rho)$, given by 
\begin{equation*}
\boldsymbol{\Sigma} _{n_0}(\rho)=\scriptsize \begin{pmatrix}
 1&  \rho &  \rho^{2}& . & . & . & \rho^{n_0-1}\\ 
 \rho&  1&  \rho&  .&  .&  .&\rho^{n_0-2} \\ 
 \rho^{2}&  \rho&  1&  .& . & . & \rho^{n_0-3}\\ 
 .&  .& . &  .&  &  & .\\ 
 .&  .&  .&  & . &  &. \\ 
 .&  .&  .&  &  & . & .\\
\rho^{n_0-1}& \rho^{n_0-2} &\rho^{n_0-3}  & . & . & . &1 \\
\end{pmatrix},   
\end{equation*}
where the correlation between adjacent elements is $\rho$, with $0<\rho<1$. Therefore, $\boldsymbol{\eta}$ is
assumed to follow $N_{10}(\boldsymbol{\mu},\boldsymbol{\Sigma}_{10}(0.3))$. The adaptive MH algorithm in Subsection \ref{ss2.3} is applied with each of the 500 simulated datasets, yielding posterior estimates $\Tilde{\theta}_{\nu};\nu=0,1,2$ and $\Tilde{\eta}_{l};l=1,\dots,10$. The results are derived from 70,000 MCMC samples, following a burn-in period of 10,000 samples and thinning to every fifteenth to mitigate autocorrelation. Convergence diagnostics of MCMC are detailed in \ref{A1}. Approximately 0.015 seconds are required for each MCMC iteration, with $n=500$.

\setlength{\tabcolsep}{3.3pt} 
\renewcommand{\arraystretch}{1.2} 
\begin{table}[h!]
\centering
\caption{
The results for regression parameters  under scenario (1) and (2) with $n=200$}
\label{frequent_fixed_scenario1_200}
\begin{tabular}{|c|c|ccccc|ccccc|}
\hline
\multirow{2}{*}{} & \multirow{2}{*}{True} & \multicolumn{5}{c|}{(1)\quad$(u_{1}, u_{2},..., u_{10})=(0.3,0.6,\dots,3.0)$}                                                                                               & \multicolumn{5}{c|}{(2)\quad$u_i\sim Uniform(0,3)$}                                                                                                \\ \cline{3-12} 
                           &                             & \multicolumn{1}{c|}{Mean}    & \multicolumn{1}{c|}{Abs. bias} & \multicolumn{1}{c|}{EPSD}    & \multicolumn{1}{c|}{SSD}    & CP   & \multicolumn{1}{c|}{Mean}    & \multicolumn{1}{c|}{Abs. bias} & \multicolumn{1}{c|}{EPSD}    & \multicolumn{1}{c|}{SSD}    & CP   \\ \hline
$\theta_{0}$                & 0.6                        & \multicolumn{1}{c|}{0.6189} & \multicolumn{1}{c|}{0.0189}        & \multicolumn{1}{c|}{0.1736} & \multicolumn{1}{c|}{0.1814} & 0.93 & \multicolumn{1}{c|}{0.5618} & \multicolumn{1}{c|}{0.0382}        & \multicolumn{1}{c|}{0.1737} & \multicolumn{1}{c|}{0.1730} & 0.95 \\
$\theta_{1}$                & -0.5                        & \multicolumn{1}{c|}{-0.5107}  & \multicolumn{1}{c|}{0.0107}        & \multicolumn{1}{c|}{0.2119} & \multicolumn{1}{c|}{ 0.2192} & 0.95 & \multicolumn{1}{c|}{-0.5167}  & \multicolumn{1}{c|}{0.0167}        & \multicolumn{1}{c|}{0.2155} & \multicolumn{1}{c|}{0.2154} & 0.97 \\
$\theta_{2}$                 & 0.7                         & \multicolumn{1}{c|}{0.7474}  & \multicolumn{1}{c|}{0.0474}        & \multicolumn{1}{c|}{0.1314} & \multicolumn{1}{c|}{0.1452} & 0.94 & \multicolumn{1}{c|}{ 0.7398 }  & \multicolumn{1}{c|}{0.0398}        & \multicolumn{1}{c|}{0.1320} & \multicolumn{1}{c|}{0.1468} & 0.92 \\
\hline
$\theta_{0}$                & -0.8                         & \multicolumn{1}{c|}{-0.8534}        & \multicolumn{1}{c|}{0.0534   }              & \multicolumn{1}{c|}{0.2231
}       & \multicolumn{1}{c|}{0.2445}       &  0.93    & \multicolumn{1}{c|}{-0.8894}        & \multicolumn{1}{c|}{0.0894}              & \multicolumn{1}{c|}{0.2297}       & \multicolumn{1}{c|}{0.2625}       &  0.93   \\
$\theta_{1}$                & -1                       & \multicolumn{1}{c|}{-1.0725}        & \multicolumn{1}{c|}{0.0725}              & \multicolumn{1}{c|}{0.3186}       & \multicolumn{1}{c|}{0.3304}       &  0.96    & \multicolumn{1}{c|}{-1.0261}        & \multicolumn{1}{c|}{0.0261}              & \multicolumn{1}{c|}{0.3224 }       & \multicolumn{1}{c|}{0.3240}       &   0.96  \\
$\theta_{2}$                 & -1.2                       & \multicolumn{1}{c|}{-1.2565}        & \multicolumn{1}{c|}{0.0565}              & \multicolumn{1}{c|}{0.1967}       & \multicolumn{1}{c|}{0.2084}       &0.96      & \multicolumn{1}{c|}{-1.2482}        & \multicolumn{1}{c|}{0.0482}              & \multicolumn{1}{c|}{0.1973}       & \multicolumn{1}{c|}{0.2132}       &  0.94    \\

\hline
$\theta_{0}$                & -0.75                        & \multicolumn{1}{c|}{-0.7929}        & \multicolumn{1}{c|}{0.0429}              & \multicolumn{1}{c|}{0.2269}       & \multicolumn{1}{c|}{0.2264}       & 0.94    & \multicolumn{1}{c|}{-0.8310}        & \multicolumn{1}{c|}{0.0810}              & \multicolumn{1}{c|}{0.2317}       & \multicolumn{1}{c|}{0.2361}       & 0.94     \\
$\theta_{1}$                & 2.1                         & \multicolumn{1}{c|}{2.2154}        & \multicolumn{1}{c|}{0.1154}              & \multicolumn{1}{c|}{0.3548}       & \multicolumn{1}{c|}{0.3955}       &0.94    & \multicolumn{1}{c|}{2.1810}        & \multicolumn{1}{c|}{0.0810}              & \multicolumn{1}{c|}{ 0.3508}       & \multicolumn{1}{c|}{0.3698}       &    0.94  \\
$\theta_{2}$                 & 1.5                        & \multicolumn{1}{c|}{1.6091}        & \multicolumn{1}{c|}{ 0.1091}              & \multicolumn{1}{c|}{0.2398}       & \multicolumn{1}{c|}{0.2744}       &  0.95    & \multicolumn{1}{c|}{1.5796}        & \multicolumn{1}{c|}{0.0796}              & \multicolumn{1}{c|}{0.2350}       & \multicolumn{1}{c|}{0.2688}       &  0.93    \\
\hline
$\theta_{0}$                & -1.5                         & \multicolumn{1}{c|}{-1.6270}        & \multicolumn{1}{c|}{0.1270}              & \multicolumn{1}{c|}{0.2964}       & \multicolumn{1}{c|}{0.3072}       &  0.96    & \multicolumn{1}{c|}{-1.6594}        & \multicolumn{1}{c|}{0.1594}              & \multicolumn{1}{c|}{ 0.3034}       & \multicolumn{1}{c|}{0.3738}       &  0.92    \\
$\theta_{1}$                & 1.7                       & \multicolumn{1}{c|}{1.8543}        & \multicolumn{1}{c|}{0.1543}              & \multicolumn{1}{c|}{0.3526}       & \multicolumn{1}{c|}{ 0.3862}       &  0.94    & \multicolumn{1}{c|}{1.7910}        & \multicolumn{1}{c|}{0.0910}              & \multicolumn{1}{c|}{0.3512}       & \multicolumn{1}{c|}{0.3883}       &   0.92   \\
$\theta_{2}$                 & -1.9                        & \multicolumn{1}{c|}{-2.0636}        & \multicolumn{1}{c|}{0.1636}              & \multicolumn{1}{c|}{0.2861}       & \multicolumn{1}{c|}{0.3221}       &0.94      & \multicolumn{1}{c|}{-2.0172}        & \multicolumn{1}{c|}{0.1172}              & \multicolumn{1}{c|}{0.2820}       & \multicolumn{1}{c|}{0.3396}       &  0.92    \\
\hline
$\theta_{0}$                & -1                          & \multicolumn{1}{c|}{-1.0782}        & \multicolumn{1}{c|}{0.0782}              & \multicolumn{1}{c|}{ 0.2543}       & \multicolumn{1}{c|}{0.2545}       &  0.96    & \multicolumn{1}{c|}{-1.0918}        & \multicolumn{1}{c|}{0.0918}              & \multicolumn{1}{c|}{0.2604}       & \multicolumn{1}{c|}{0.2993}       &  0.92    \\
$\theta_{1}$                & -1.25                       & \multicolumn{1}{c|}{-1.3345}        & \multicolumn{1}{c|}{0.0845}              & \multicolumn{1}{c|}{0.3592}       & \multicolumn{1}{c|}{0.4029}       &  0.92   & \multicolumn{1}{c|}{-1.3303}        & \multicolumn{1}{c|}{0.0803}              & \multicolumn{1}{c|}{ 0.3609}       & \multicolumn{1}{c|}{0.3762}       &  0.97   \\
$\theta_{2}$                 & 1.75                         & \multicolumn{1}{c|}{1.8775}        & \multicolumn{1}{c|}{0.1275}              & \multicolumn{1}{c|}{0.2717}       & \multicolumn{1}{c|}{0.3251}       &    0.92 & \multicolumn{1}{c|}{1.8453}        & \multicolumn{1}{c|}{0.0953}              & \multicolumn{1}{c|}{0.2688}       & \multicolumn{1}{c|}{0.3094}       &  0.94    \\
\hline

\end{tabular}
\end{table}

\setlength{\tabcolsep}{0.7pt} 
\renewcommand{\arraystretch}{1.2} 

\setlength{\tabcolsep}{3.3pt} 
\renewcommand{\arraystretch}{1.2} 
\begin{table}[h!]
\centering
\caption{
The results for regression parameters  under scenario (1) and (2) with $n=500$}
\label{frequent_fixed_scenario1_500}
\begin{tabular}{|c|c|ccccc|ccccc|}
\hline
\multirow{2}{*}{} & \multirow{2}{*}{True} & \multicolumn{5}{c|}{(1)\quad$(u_{1}, u_{2},..., u_{10})=(0.3,0.6,\dots,3.0)$}                                                                                               & \multicolumn{5}{c|}{(2)\quad$u_i\sim Uniform(0,3)$}                                                                                                \\ \cline{3-12} 
                           &                             & \multicolumn{1}{c|}{Mean}    & \multicolumn{1}{c|}{Abs. bias} & \multicolumn{1}{c|}{EPSD}    & \multicolumn{1}{c|}{SSD}    & CP   & \multicolumn{1}{c|}{Mean}    & \multicolumn{1}{c|}{Abs. bias} & \multicolumn{1}{c|}{EPSD}    & \multicolumn{1}{c|}{SSD}    & CP   \\ \hline
$\theta_{0}$                & 0.6                        & \multicolumn{1}{c|}{0.5971} & \multicolumn{1}{c|}{0.0029}        & \multicolumn{1}{c|}{0.1104} & \multicolumn{1}{c|}{0.0975} & 0.98 & \multicolumn{1}{c|}{0.5374} & \multicolumn{1}{c|}{0.0266}        & \multicolumn{1}{c|}{0.1132} & \multicolumn{1}{c|}{0.1111} & 0.95 \\
$\theta_{1}$                & -0.5                        & \multicolumn{1}{c|}{-0.5064}  & \multicolumn{1}{c|}{0.0064}        & \multicolumn{1}{c|}{0.1314} & \multicolumn{1}{c|}{ 0.1382} & 0.94 & \multicolumn{1}{c|}{-0.5103}  & \multicolumn{1}{c|}{0.0103}        & \multicolumn{1}{c|}{0.1335} & \multicolumn{1}{c|}{0.1387} & 0.94 \\
$\theta_{2}$                 & 0.7                         & \multicolumn{1}{c|}{0.7161}  & \multicolumn{1}{c|}{0.0161}        & \multicolumn{1}{c|}{0.0804} & \multicolumn{1}{c|}{0.0860} & 0.93 & \multicolumn{1}{c|}{ 0.6978 }  & \multicolumn{1}{c|}{0.0022}        & \multicolumn{1}{c|}{0.0803} & \multicolumn{1}{c|}{0.0783} & 0.96 \\
\hline
$\theta_{0}$                & -0.8                         & \multicolumn{1}{c|}{-0.8056}        & \multicolumn{1}{c|}{0.0056   }              & \multicolumn{1}{c|}{0.1440
}       & \multicolumn{1}{c|}{0.1403}       &  0.96    & \multicolumn{1}{c|}{-0.8628}        & \multicolumn{1}{c|}{0.0628}              & \multicolumn{1}{c|}{0.1501}       & \multicolumn{1}{c|}{0.1660}       &  0.93   \\
$\theta_{1}$                & -1                       & \multicolumn{1}{c|}{-1.0423}        & \multicolumn{1}{c|}{0.0423}              & \multicolumn{1}{c|}{0.1948}       & \multicolumn{1}{c|}{0.2198}       &  0.92    & \multicolumn{1}{c|}{-1.0165}        & \multicolumn{1}{c|}{0.0165}              & \multicolumn{1}{c|}{0.1999 }       & \multicolumn{1}{c|}{0.1980}       &   0.94  \\
$\theta_{2}$                 & -1.2                       & \multicolumn{1}{c|}{-1.2288}        & \multicolumn{1}{c|}{0.0288}              & \multicolumn{1}{c|}{0.1192}       & \multicolumn{1}{c|}{0.1211}       &0.96      & \multicolumn{1}{c|}{-1.2354}        & \multicolumn{1}{c|}{0.0354}              & \multicolumn{1}{c|}{0.1227}       & \multicolumn{1}{c|}{0.1223}       &  0.96    \\

\hline
$\theta_{0}$                & -0.75                        & \multicolumn{1}{c|}{-0.7820}        & \multicolumn{1}{c|}{0.0320}              & \multicolumn{1}{c|}{0.1479}       & \multicolumn{1}{c|}{0.1476}       & 0.97    & \multicolumn{1}{c|}{-0.7926}        & \multicolumn{1}{c|}{0.0426}              & \multicolumn{1}{c|}{0.1494}       & \multicolumn{1}{c|}{0.1735}       & 0.93     \\
$\theta_{1}$                & 2.1                         & \multicolumn{1}{c|}{2.1455}        & \multicolumn{1}{c|}{0.0455}              & \multicolumn{1}{c|}{0.2156}       & \multicolumn{1}{c|}{0.2295}       &0.96      & \multicolumn{1}{c|}{2.0954}        & \multicolumn{1}{c|}{0.0046}              & \multicolumn{1}{c|}{ 0.2154}       & \multicolumn{1}{c|}{0.2274}       &    0.94  \\
$\theta_{2}$                 & 1.5                        & \multicolumn{1}{c|}{1.5416}        & \multicolumn{1}{c|}{ 0.0416}              & \multicolumn{1}{c|}{0.1447}       & \multicolumn{1}{c|}{0.1714}       &  0.92    & \multicolumn{1}{c|}{1.5244}        & \multicolumn{1}{c|}{0.0244}              & \multicolumn{1}{c|}{0.1439}       & \multicolumn{1}{c|}{0.1537}       &  0.94    \\
\hline
$\theta_{0}$                & -1.5                         & \multicolumn{1}{c|}{-1.5561}        & \multicolumn{1}{c|}{0.0561}              & \multicolumn{1}{c|}{0.1851}       & \multicolumn{1}{c|}{0.1989}       &  0.94    & \multicolumn{1}{c|}{-1.5589}        & \multicolumn{1}{c|}{0.0589}              & \multicolumn{1}{c|}{ 0.1869}       & \multicolumn{1}{c|}{0.2024}       &  0.92    \\
$\theta_{1}$                & 1.7                       & \multicolumn{1}{c|}{1.7603}        & \multicolumn{1}{c|}{0.0603}              & \multicolumn{1}{c|}{0.2137}       & \multicolumn{1}{c|}{ 0.2317}       &  0.94    & \multicolumn{1}{c|}{1.7262}        & \multicolumn{1}{c|}{0.0262}              & \multicolumn{1}{c|}{0.2141}       & \multicolumn{1}{c|}{0.2124}       &   0.94   \\
$\theta_{2}$                 & -1.9                        & \multicolumn{1}{c|}{-1.9790}        & \multicolumn{1}{c|}{0.0790}              & \multicolumn{1}{c|}{0.1725}       & \multicolumn{1}{c|}{0.2045}       &0.92      & \multicolumn{1}{c|}{-1.9332}        & \multicolumn{1}{c|}{0.0332}              & \multicolumn{1}{c|}{0.1699}       & \multicolumn{1}{c|}{0.1925}       &  0.94    \\
\hline
$\theta_{0}$                & -1                          & \multicolumn{1}{c|}{-1.0200}        & \multicolumn{1}{c|}{0.0200}              & \multicolumn{1}{c|}{ 0.1632}       & \multicolumn{1}{c|}{0.1517}       &  0.93    & \multicolumn{1}{c|}{-1.0416}        & \multicolumn{1}{c|}{0.0416}              & \multicolumn{1}{c|}{0.1679}       & \multicolumn{1}{c|}{0.1703}       &  0.96    \\
$\theta_{1}$                & -1.25                       & \multicolumn{1}{c|}{-1.2962}        & \multicolumn{1}{c|}{0.0462}              & \multicolumn{1}{c|}{0.2177}       & \multicolumn{1}{c|}{0.2386}       &  0.93   & \multicolumn{1}{c|}{-1.2726}        & \multicolumn{1}{c|}{0.0226}              & \multicolumn{1}{c|}{ 0.2228}       & \multicolumn{1}{c|}{0.2339}       &  0.94    \\
$\theta_{2}$                 & 1.75                         & \multicolumn{1}{c|}{1.7898}        & \multicolumn{1}{c|}{0.0398}              & \multicolumn{1}{c|}{0.1616}       & \multicolumn{1}{c|}{0.1612}       &    0.95 & \multicolumn{1}{c|}{1.7668}        & \multicolumn{1}{c|}{0.0168}              & \multicolumn{1}{c|}{0.1621}       & \multicolumn{1}{c|}{0.1756}       &  0.94    \\
\hline
\end{tabular}
\end{table}
Tables \ref{frequent_fixed_scenario1_200}  and \ref{frequent_fixed_scenario1_500} present the frequentist operating characteristics of $\Tilde{\theta}_{0}$, $\Tilde{\theta}_{1}$,  and  $\Tilde{\theta}_{2}$ under scenarios $(1)\,(s_{1}, \dots, s_{10})=(0.3,\dots,3)$ and $(2)\,s_{i}\sim U(0,3)$, for $n=200$ and $n=500$ respectively. Here, `Mean' is the average of 500 posterior mean estimates. Absolute value of bias is denoted by Abs. bias. The standard deviation of every posterior sample is calculated and these estimated posterior standard deviations are averaged across all replications to generate EPSD. Additionally, SSD is determined by computing the sample standard deviation of the posterior mean estimates. Percentile approach is opted to form 95\% Bayesian credible intervals (BCI) and the percentage of these BCIs encompassing the true value is given by CP, the coverage probability. The efficacy of the proposed method under both fixed and random censoring schemes for both the sample sizes is evident as the mean values closely approximate the true values of  $\theta_{0}$, $\theta_{1}$,  and  $\theta_{2}$, EPSD and SSD exhibit minimal differences as well as low magnitudes, and CP values are close to 0.95. With increase in sample size, reduction of Abs. bias, EPSD and SSD is also observed.

\begin{table}[h!]
\centering
\caption{
The maximum value among the local MSEs of $\Tilde{F_{s}}(t)$ under scenario (1) and (2) with $n=200$}
\label{maxmse_200}
\small{
\begin{tabular}{|c|c|c|c|c|c|c|c|}
\hline
                                               & (0.6,-0.5,0.7) & (-0.8,-1,-1.2)  & (-0.75,2.1,1.5)& (-1.5,1.7,-1.9) & (-1,-1.25,1.75) \\ \hline
$(1)$ & 0.1712        & 0.1713        & 0.2072          & 0.1814         & 0.1961            \\ 
$(2)$                         & 0.2391         & 0.3281        & 0.2596       & 0.3012         & 0.2181                 \\ \hline
\end{tabular}}
\end{table}

\begin{table}[h!]
\centering
\caption{
The maximum value among the local MSEs of $\Tilde{F_{s}}(t)$ under scenario (1) and (2) with $n=500$}
\label{maxmse_500}
\small{
\begin{tabular}{|c|c|c|c|c|c|c|c|}
\hline
                                               & (0.6,-0.5,0.7) & (-0.8,-1,-1.2)  & (-0.75,2.1,1.5)& (-1.5,1.7,-1.9) & (-1,-1.25,1.75) \\ \hline
$(1)$ & 0.0906        & 0.1399         & 0.1574          & 0.1178         & 0.1419             \\ 
$(2)$                         & 0.1203         & 0.1939        & 0.1301         & 0.1868          & 0.2073                 \\ \hline
\end{tabular}}
\end{table}

\begin{figure}[h!]
\centering
\includegraphics[width=15cm,height=8cm]{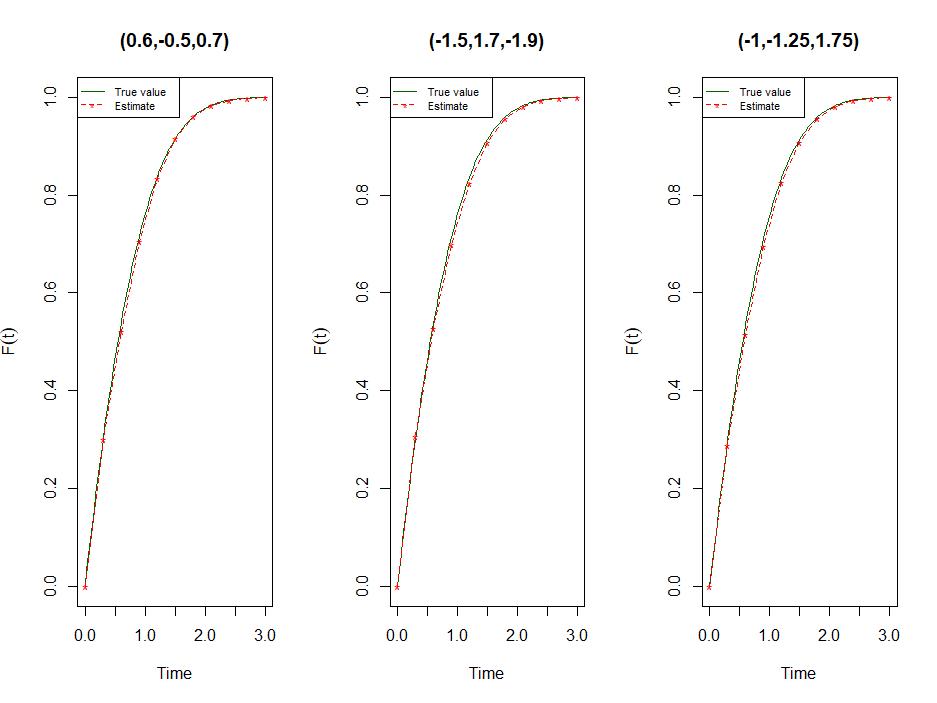}
\caption{The estimates of $F(t)$ alongside the true curve}
\label{Fig_comparison}
\end{figure}

Using estimates of $\eta_{l}; l=1,\dots,10$, obtained from five different settings, $\Tilde{F}_{s}(t)$ are computed by \eqref{2.6} at these ten distinct monitoring times. The effectiveness of the proposed method in estimating distribution function is evaluated by computing the local mean square errors of $\Tilde{F}_{s}(t)$ at 10 distinct monitoring times. Tables \ref{maxmse_200} and \ref{maxmse_500} summarise the maximum of these errors (MaxMSE) for $n=200$ and $n=500$ respectively, which are consistently minimal across all setups and smaller for larger samples, indicating accurate estimation of $F(t)$. Additionally, Fig. \ref{Fig_comparison} illustrates the comparison between the actual $F(t)$ curves and the estimated curves (constructed by connecting the estimates at ten specific monitoring times, allowing for a direct comparison against the true $F(t)$ curves) for three setups, demonstrating the consistent performance of the proposed estimation procedure.

\section{Illustrative Examples}\label{sec4}

\subsection{Lung Tumor Data}
\cite{sun2006statistical} has presented the data from a tumorigenicity experiment conducted by \cite{hoel1972representation} involving 144 RFM mice. The study is designed to evaluate the influence of environmental conditions on the time to lung tumor onset ($T_i$) in mice. As tumor onset is not directly observable, only the time of death or sacrifice ($U_i$) and an indicator of the presence or absence of a tumor at that time ($\delta_i$) are recorded, resulting in current status data. Since lung tumors in RFM mice are non-lethal, tumor onset has not affected their mortality and the event time $T_i$ remains independent of the examination time $U_i$.

\begin{figure}[h!]
\centering
\includegraphics[width=7cm,height=4cm]{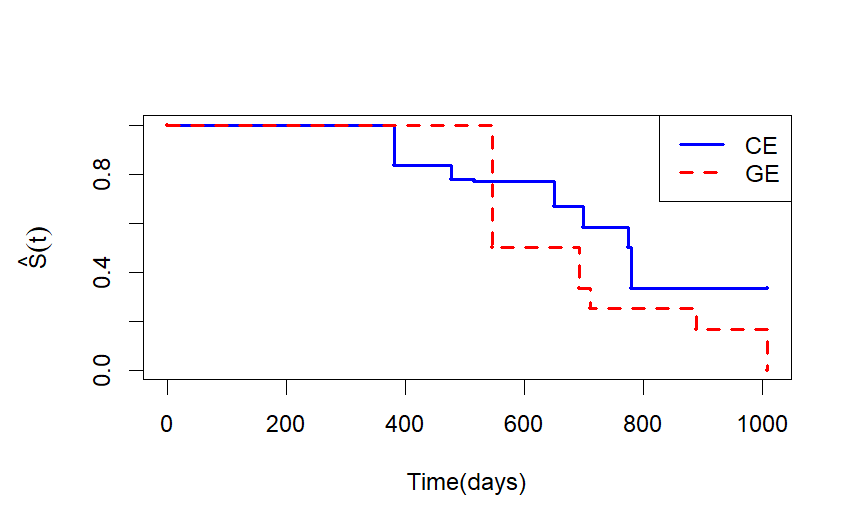}
\caption{
NPMLE of $S(t)$ for lung tumor data }
\label{KM_tumor}
\end{figure}

The experiment involved placing 96 mice in a conventional environment (CE) and 48 in a germ-free environment (GE). The median follow-up time is 662.5 days and the event times are censored for 82 mice (71.9\% in CE, 27.1\% in GE). To assess the presence of a cured fraction, the non-parametric maximum likelihood estimator (NPMLE) of the survival function, $\hat{S}(\cdot)$, was plotted for both environments, as shown in Fig. \ref{KM_tumor}. A long plateau at the tail of $\hat{S}(\cdot)$ for CE indicates the presence of a cure fraction among the mice in the conventional environment. According to \cite{maller1996survival}, one may look for fitting a cure model to the data in order to estimate the impact of environment on the time to tumor onset  and the cure fraction.

In order to begin with the analysis, the data is modified by selecting eleven distinct monitoring times $(s_1,\dots,s_{11})=(45,381,477,515,650,679,773,779,839,888,1008)$ each representing the time intervals during which $\hat{S}(\cdot)$ remains piecewise constant. As an informative prior for $\boldsymbol{\eta}$, 
$N_{11}(\boldsymbol{\mu},0.1*\boldsymbol{\Sigma}_{11}(0.3))$ is chosen, where $\mu_l=\log\left(-\log\left(\frac{\hat{S}(s_l)}{\hat{S}(s_{l-1})}\right) \right)$ for $l=1,\dots,11$. \cite{wang2020efficient} has introduced a binary covariate $X$ to represent the environment, assigning 0 for CE and 1 for GE, and has applied maximum likelihood estimation to fit equation \eqref{1.2} to the data. Building on this, informative normal priors centered at their MLEs with minimum spread are used for the regression coefficients: $\theta_0\sim N(-0.27,0.01^2)$ and $\theta_1\sim N(0.81,0.01^2)$. Posterior computation is performed with the adaptive MH algorithm (see Subsection \ref{ss2.3}) and the results are reported in Table \ref{LTsumm}, with MCMC diagnostics detailed in \ref{A2}.

\setlength{\tabcolsep}{5pt} 
\renewcommand{\arraystretch}{1.0} 
\begin{table}[h]
\centering
\caption{
Summary of Bayesian estimates for the lung tumor data }
\label{LTsumm}
\begin{tabular}{|c|c|c|c|c|}
\hline
Parameters        & Estimates & Posterior standard deviations & BCI \\ \hline
$\theta_{0}$ &  -0.2702  &  0.0100       & (-0.2906,-0.2502)              \\
$\theta_{1}$ &   0.8102 &  0.0102      & (0.7899, 0.8298)           \\
 \hline      
\end{tabular}
\end{table}

\begin{figure}[h!]
\centering
\includegraphics[width=7cm,height=4cm]{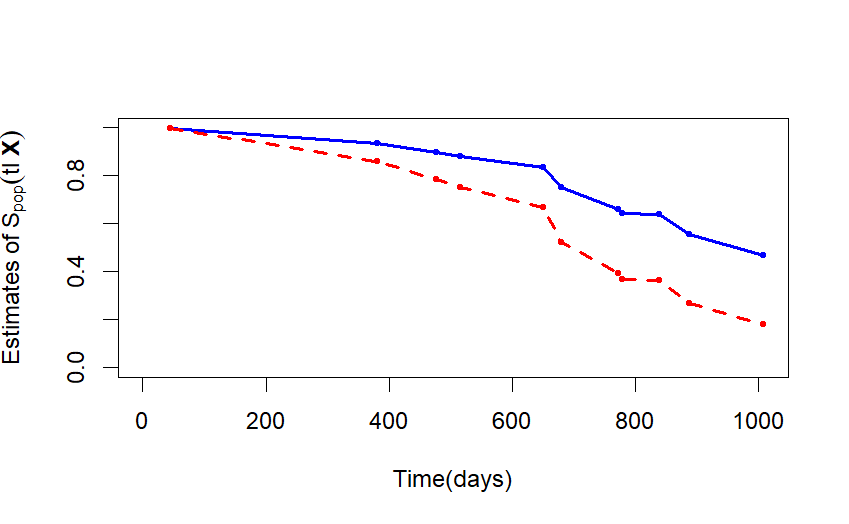}
\caption{
Estimates of population survival function for lung tumor data }
\label{surv_lungtum}
\end{figure}

\begin{figure}[h!]
\centering
\includegraphics[width=7cm,height=4cm]{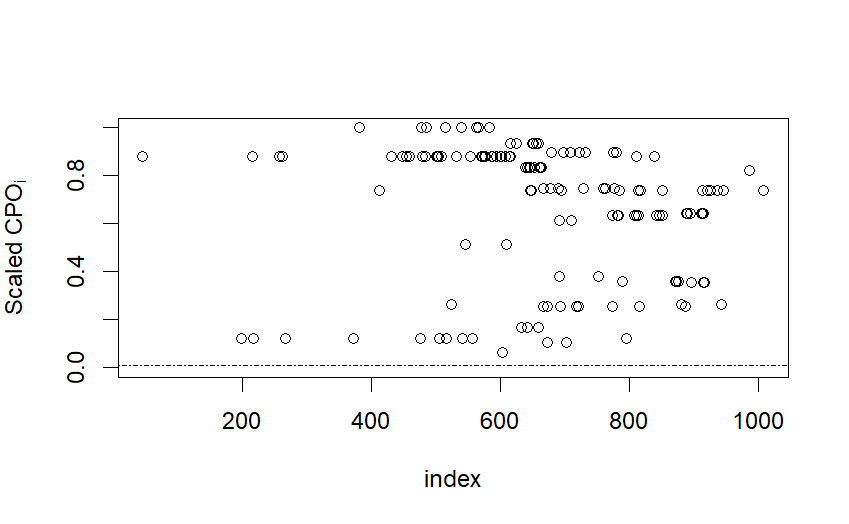}
\caption{
Scaled CPO plot for lung tumor data}
\label{cpo_lung}
\end{figure}

The estimates for $\theta_0$ (negative) and $\theta_1$ (positive) are statistically significant, as their BCIs exclude zero. This indicates that the environment significantly affects lung tumor risk, with mice in the germ-free environment (GE) being at higher risk. Using  \eqref{2.8}, the cure rates are estimated as 0.4662 for CE and 0.1798 for GE, indicating that mice in CE are more likely to be cured. The population survival curves, estimated using \eqref{2.7} and shown in Fig. \ref{surv_lungtum} (constructed by connecting the estimates at distinct observational times), further confirm that survival probabilities are higher for mice in CE compared to GE. 
Fig. \ref{cpo_lung} shows scaled CPOs plotted against the mice indices, with no visible pattern or outliers, suggesting a good model fit. The LPML and DIC of the model are -87.96 and 91.08 respectively.  

\subsection{Breast Cancer Data}
\cite{losch1998prognostic} gathered data from 100 women with primary invasive ductal carcinoma, who underwent initial surgical treatment, followed by monitoring for the first confirmed metastasis or recurrence. This breast cancer dataset utilised by \cite{heinze2001solution} and now accessible in the \textit{coxphf} package of \textit{R} software, provides information on the recurrence-free interval, representing the duration between the initial surgical procedure and the first verified metastasis or recurrence, subject to right censoring.
Additionally, it incorporates four potential risk factors: $X_1$=tumor stage (i.e. 1, if stage is 2, 3 or 4 and 0, if
stage is 1),  $X_2$=histological grading (i.e. 1, if grade is 2 or 3 and 0, if
grade is 1), $X_3$=nodal status (i.e. 1, if number of nodes is 1 or 2 and 0, if
number of nodes is 0), and $X_4$=cathepsin D (CD) immunoreactivity (i.e. 1, if CD positive and 0, if CD negative), intended for evaluating their impact on survival time.

\begin{figure}[h!]
\centering
\includegraphics[width=9cm,height=4cm]{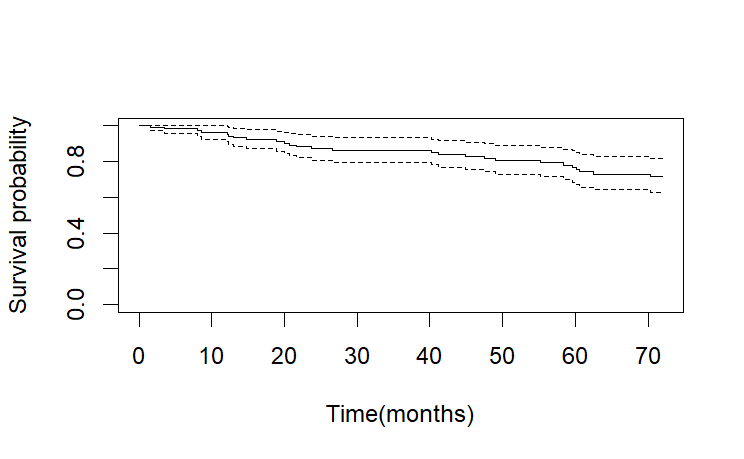}
\caption{
Kaplan-Meier Survival curve of breast cancer data }
\label{KM_breast}
\end{figure}

The median follow-up time for the patients is 72 months. Among the 100 patients, 74\% are censored, with 51\% of the censoring occurring at the maximum follow-up time of 72 months. According to \cite{maller1996survival}, The presence of several censored observations near the largest monitoring time and the Kaplan-Meier estimator's plateau around 0.714 in Fig. \ref{KM_breast} indicate the potential presence of a cure fraction and the suitability of a cure model. To illustrate the proposed Bayesian estimation procedure, the data are transformed into current status data, by grouping lifetime data into 12 non-overlapping intervals ([0,6), [6,12), [12,18), [18,24), [24,30), [30,36), [36,42), [42,48), [48,54), [54,60), [60,66), and [66,72]) and converting these into exact observational times by considering their midpoints, simplifying the implementation.

Unlike the previous example, a promotion time cure model has not yet been fitted to the data to avail some prior information regarding the parameter values. Therefore, careful prior elicitation is crucial to reduce uncertainty around the parameters and ensure meaningful posterior probabilities. To determine the priors for regression coefficients, cure probabilities are obtained using the Kaplan-Meier curve of the survival function and a method for prior elicitation outlined in \cite{lambert2019estimation} is implemented. This resulted in the informative priors, $\theta_{0}\sim N(-1.171, 0.212^{2})$, $\theta_{1}\sim N(0.737,0.044^{2})$, $\theta_{2}\sim N(0.378,0.002^{2})$,  $\theta_{3}\sim N(0.789,0.077^{2})$, and $\theta_{4}\sim N(0.587,0.095^{2})$. In order to obtain a prior for $\boldsymbol{\eta}$, the Kaplan-Meier estimator of the baseline survival function at distinct monitoring times $s_l;l=1,\dots,12$  are noted as $\hat{S}_{pop}(s_l|\textbf{X}=0)$ and an estimator for $F(.)$ at $s_l$ is obtained using $\hat{F}(s_l)=\frac{-\log \hat{S}_{pop}(t|\textbf{X}=0)}{mean(\theta_0)}$ for $l=1,\dots,12$. Further, $N_{12}(\boldsymbol{\mu},\boldsymbol{\Sigma}_{12}(0.3))$ is adopted as $\boldsymbol{\eta}$-prior, with $\mu_l=\log\left(-\log\left(\frac{1-\hat{F}(s_l)}{1-\hat{F}(s_{l-1})}\right) \right)$ for $l=1,\dots,12$.  Posterior summary of the parameter estimation on implementing the proposed adaptive MH algorithm is shown in Table \ref{BCsumm}. One can see \ref{A3} for more details on the computation as well as Markov chain diagnostics.

\setlength{\tabcolsep}{5pt} 
\renewcommand{\arraystretch}{1.0} 
\begin{table}[h]
\centering
\caption{
Summary of Bayesian estimates with the breast cancer data }
\label{BCsumm}
\begin{tabular}{|c|c|c|c|c|}
\hline
Parameters        & Estimates & Posterior standard deviations & BCI \\ \hline
$\theta_{0}$ &  -1.7986  &  0.1366       & (-2.0721,-1.5332)              \\
$\theta_{1}$ &   0.7311  &  0.0459       & (0.6402, 0.8202)           \\
$\theta_{2}$              & 0.3778   &  0.0025    & (0.3729,0.3829)         \\
$\theta_{3}$              & 0.7718  & 0.0753       & (0.6315,0.9250)  \\  
$\theta_{4}$              & 0.5553  & 0.0896      & (0.3756,0.7252)  \\  \hline       
\end{tabular}
\end{table}

 The estimates of all regression coefficients except intercept are positive and statistically significant, as none of their BCIs include zero.  Using these estimates, the cure rates for patients at various levels of covariates $X_{j};j=1,2,3,4$ can be estimated using \eqref{2.8}. For instance, an individual diagnosed with tumor stage 1 and being CD positive has a cure rate of 0.5559, which is lower than the baseline cure rate of 0.8474, estimated by setting all the covariates at level zero. Additionally, one can observe that the cure rates at unfavorable levels of covariates are lower than the baseline cure rate. These observations affirm that the patients in higher stage of tumor, higher levels of grading, having higher number of nodes, and CD positivity are less likely to cure the risk of tumor recurrence in comparison with others. 
 
\begin{figure}[h!]
\centering
\includegraphics[width=9cm,height=3cm]
{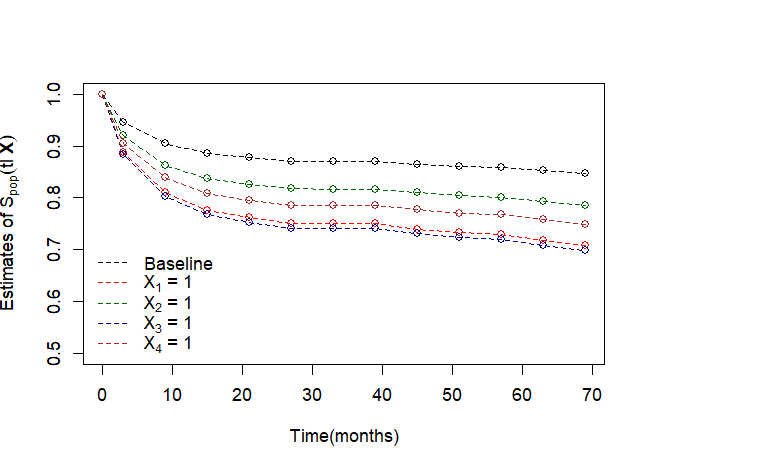}
\caption{
Estimates of population survival function for breast cancer data 
}
\label{BCfig}
\end{figure}

In addition to the cure rate, the population survival functions are estimated as shown in Fig. \ref{BCfig}, demonstrating that unfavorable levels of covariates are associated with lower survival probabilities. Therefore, all four risk factors are recognised as prognostic markers in the context of breast cancer, with higher tumor stage and grading, increased number of involved lymph nodes, and positivity for Cathepsin D pertaining to a significantly negative prognostic impact on survival. This information enhances our comprehension of the biological attributes of breast cancer and aids in patient management and treatment decision-making. 

\begin{figure}[h!]
\centering
\includegraphics[width=7cm,height=3cm]{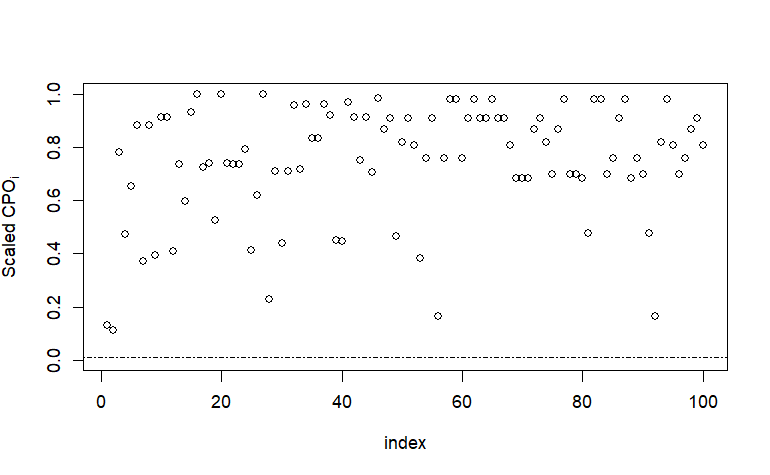}
\caption{
Scaled CPO plot for breast cancer data }
\label{cpo_bc}
\end{figure}

Furthermore, the plot of the Bayesian estimates of the baseline survival function in Fig. \ref{BCfig} closely matches the Kaplan-Meier curve for actual data in Fig. \ref{KM_breast}. The random pattern of scaled CPOs in Fig. \ref{cpo_bc} suggests that the model fits to the data adequately. The model's LPML and DIC are -50.57 and 99.68 respectively.

\section{Conclusions}\label{sec5}
The paper has presented a comprehensive framework for a Bayesian promotion time cure model, designed to analyse current status data. The Bayesian estimation procedure adopted in this study is notable for its utilisation of proper priors for the model parameters and an adaptive Metropolis-Hastings algorithm for posterior computation. This model not only advances the existing methods of estimation, but also introduces novel techniques for model comparison and validation using cross-validated predictive ordinate (CPO).  The introduced model contributes significantly to the Bayesian cure model literature, as established through simulation results and practical data analyses. By incorporating prior knowledge and uncertainty, this approach yields more reliable parameter estimates, thereby enhancing the overall robustness and applicability of cure modelling techniques.

In the literature, Box-Cox-type transformations including specific cases like the proportional hazards and proportional odds cure models are used in cure rate models to enhance model flexibility. Bayesian techniques can be devised for these models to offer robust parameter estimation and uncertainty quantification. The current status censoring becomes informative when time of examination is associated with the time of event. To address this issue and prevent potential bias, a shared frailty model can be employed, capturing the interplay between event occurrence time and time of examination. Our intention is to further investigate Bayesian extensions in this direction, aiming to refine our understanding of complex disease dynamics and improve model performance in clinical research.


\section*{Acknowledgements}
The first author wishes to acknowledge the financial support of the Council of Scientific \& Industrial Research, Government of India,  via the Junior Research Fellowship scheme under reference No. 09/0239(13499)/2022-EMR-I.

\bibliographystyle{apalike}
\bibliography{ref}

\section*{Conflicts of Interest}
There are no conflicts of interest between the authors.

\appendix
\counterwithout{figure}{section}

\section{MCMC  Convergence and Mixing Diagnostics}\label{A}
Convergence of Markov chains for simulation studies and real data analyses is demonstrated within the Appendix. As graphical checks, autocorrelation plots (ACF plots), trace plots, and posterior histograms are examined. Gelman-Rubin diagnostics, effective sample sizes (ESS), and acceptance rate are also reported.

\subsection{Simulation Studies}\label{A1}
\begin{figure}[h!]     
     \begin{subfigure}[b]{0.32\textwidth}
    \includegraphics[width=\textwidth]{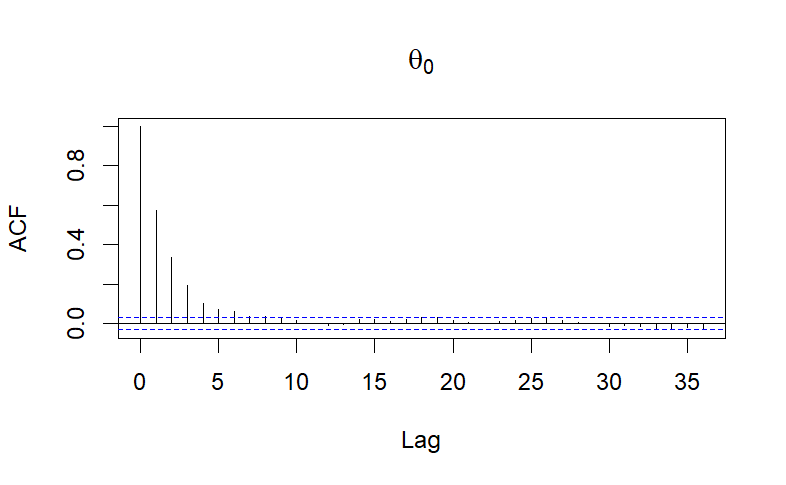}
     \end{subfigure}
     \hfill
     \begin{subfigure}[b]{0.32\textwidth}
         \includegraphics[width=\textwidth]{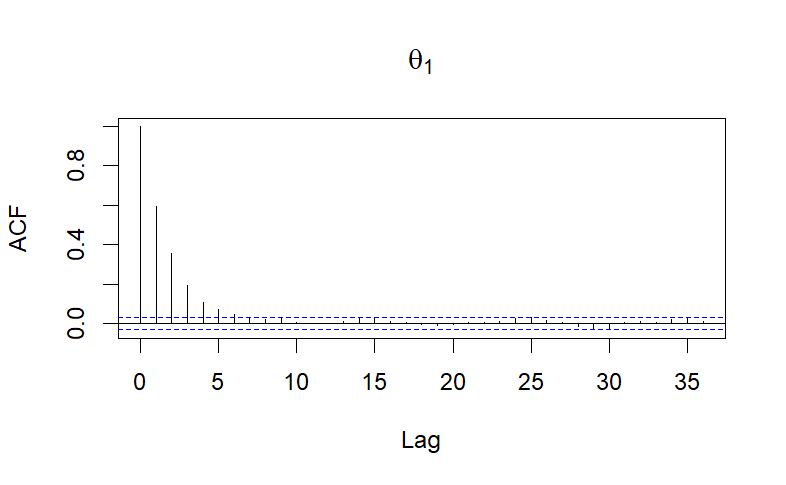}
     \end{subfigure}
     \hfill
     \begin{subfigure}[b]{0.32\textwidth}
         \includegraphics[width=\textwidth]{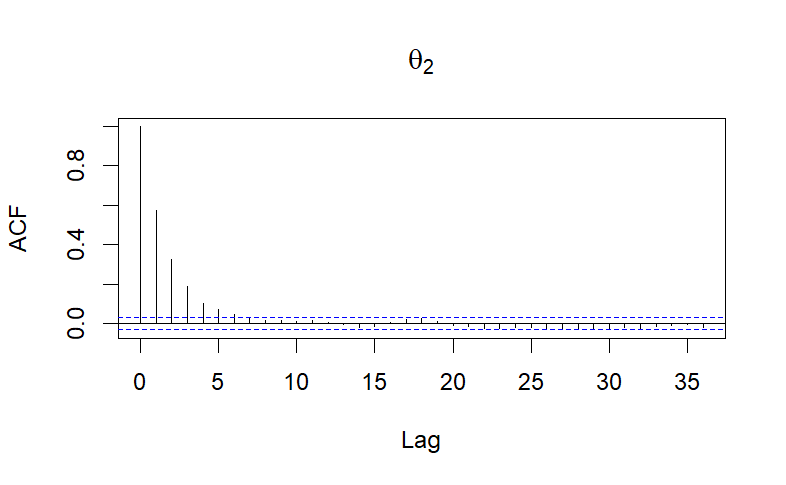}
     \end{subfigure}
     \caption{ACF plots of parameters when ($\theta_0,\theta_1,\theta_2)=(0.6,-0.5,0.7)$} 
        \label{acf_sim}
\end{figure}

\begin{figure}[h!]
     \begin{subfigure}[b]{0.32\textwidth}
    \includegraphics[width=\textwidth]{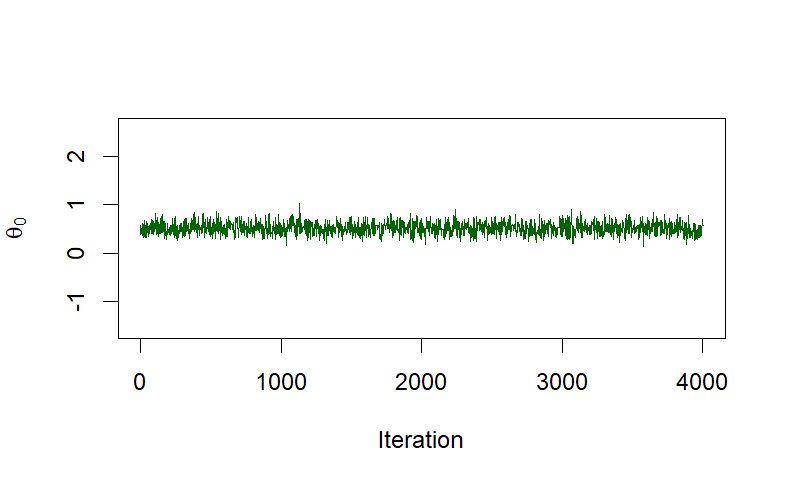}
     \end{subfigure}
     \hfill
     \begin{subfigure}[b]{0.32\textwidth}
         \includegraphics[width=\textwidth]{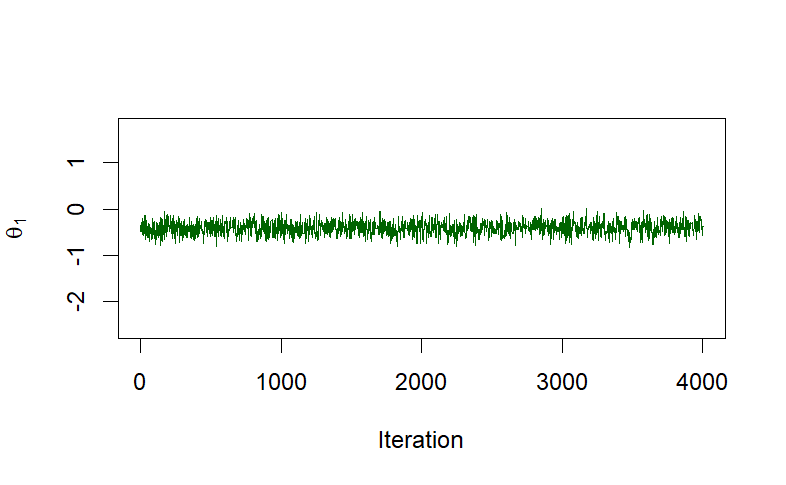}
     \end{subfigure}
     \hfill
     \begin{subfigure}[b]{0.32\textwidth}
         \includegraphics[width=\textwidth]{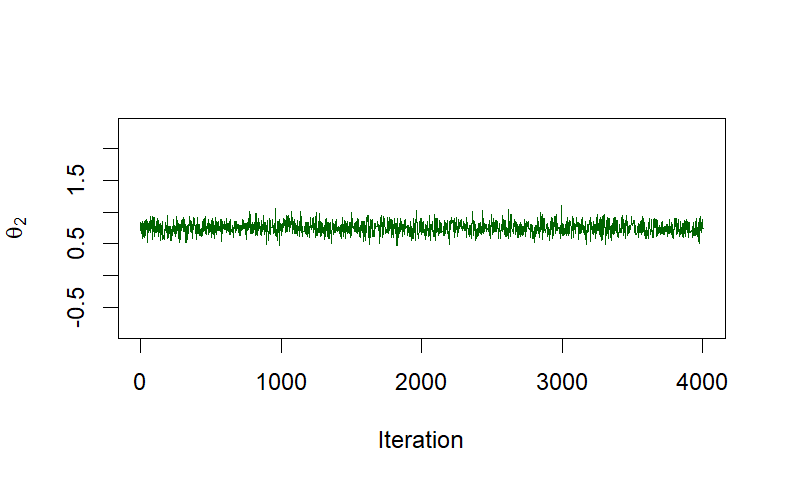}
     \end{subfigure}
     \caption{Trace plots of parameters when ($\theta_0,\theta_1,\theta_2)=(0.6,-0.5,0.7)$}    \label{trace_sim}
\end{figure}

\begin{figure}[h!]
     \begin{subfigure}[b]{0.32\textwidth}
    \includegraphics[width=\textwidth]{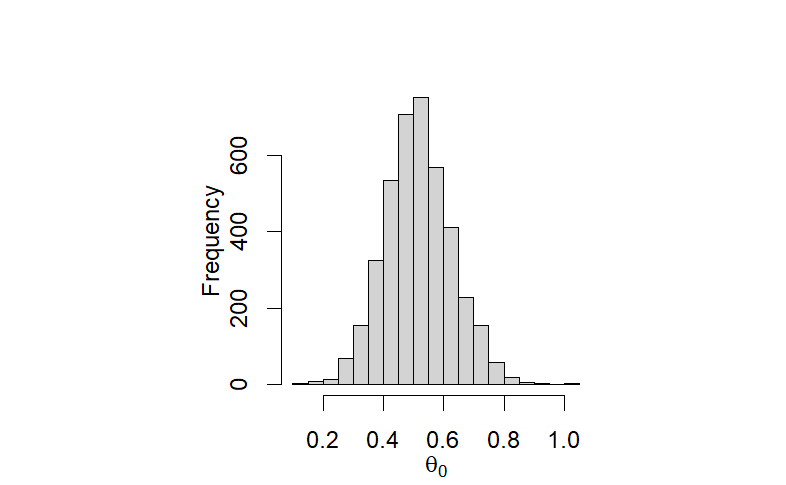}
     \end{subfigure}
     \hfill
     \begin{subfigure}[b]{0.32\textwidth}
         \includegraphics[width=\textwidth]{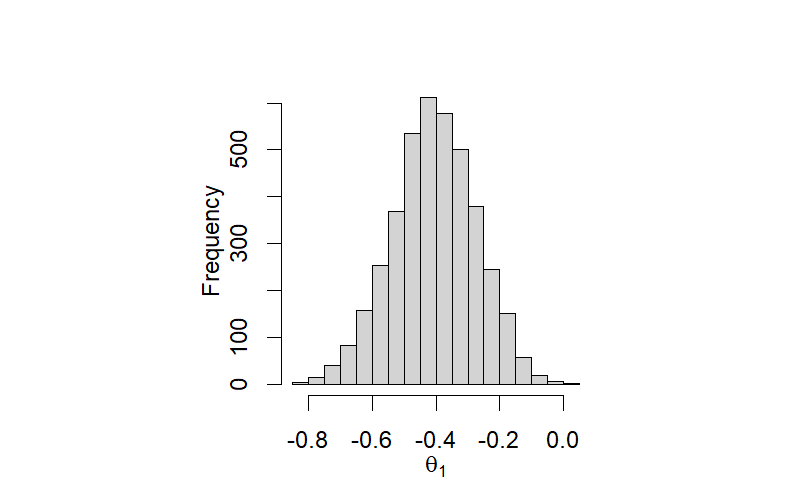}
     \end{subfigure}
     \hfill
     \begin{subfigure}[b]{0.32\textwidth}
         \includegraphics[width=\textwidth]{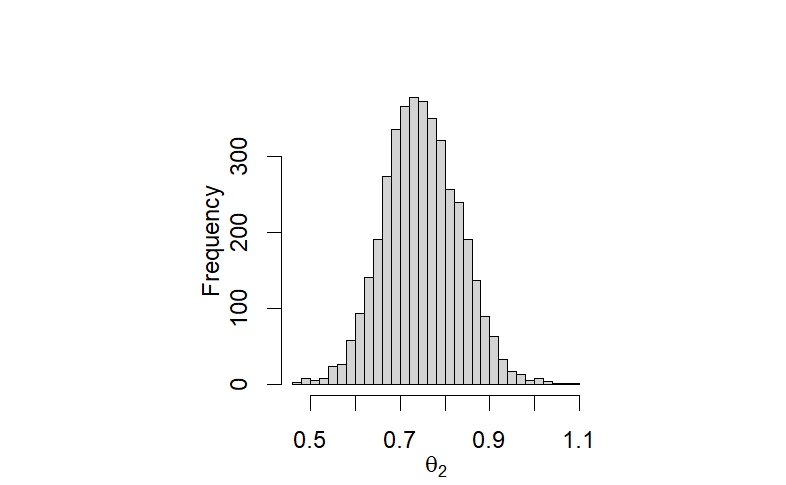}
     \end{subfigure}
     \caption{Posterior histograms of parameters when ($\theta_0,\theta_1,\theta_2)=(0.6,-0.5,0.7)$}    \label{pd_sim}
\end{figure}

Consider $(\theta_{0},\theta_{1},\theta_{2})=(0.6,-0.5,0.7)$ under scenario (1).
With a randomly generated data, 70,000 MCMC simulations are done. As burn-in, 10,000 samples are removed and the remaining ones are thinned, keeping only multiples of 15. ACF plots in Fig. \ref{acf_sim} visualise chain autocorrelation: high autocorrelation implies poor mixing, while low values indicate better convergence. The plotted parameters show rapid autocorrelation decay, indicating well-behaved simulated Markov chains. Trace plots in Fig. \ref{trace_sim} depict the evolution of MCMC-generated samples for parameters of interest. The plots show random fluctuations around a central value without anomalies, suggesting well-mixed posterior samples. Posterior histograms in Fig. \ref{pd_sim} are stable with consistent shapes and narrow peaks. This indicates convergence and low uncertainty in parameter estimation. Gelman-Rubin diagnostics assesses convergence across multiple chains. Values of potential scale reduction factor close to 1 for $\theta_{0}$, $
\theta_{1}$, and $\theta_{2}$ indicate convergence, as observed with ten independent chains. The ESS for $\theta_{0}$, $\theta_{1}$, and  $\theta_{2}$ are 618, 616, and 673 respectively. Every MCMC repetition requires approximately 0.015 seconds. The rate of acceptance is 0.0969.

\subsection{Lung Tumor Data Analysis }\label{A2}
\begin{figure}[h!]
\centering
\includegraphics[width=.4\textwidth]{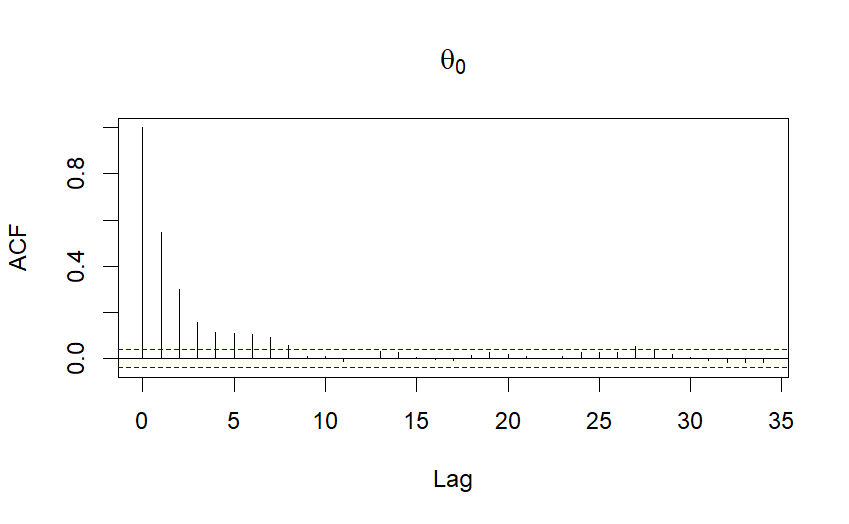}\quad
\includegraphics[width=.4\textwidth]{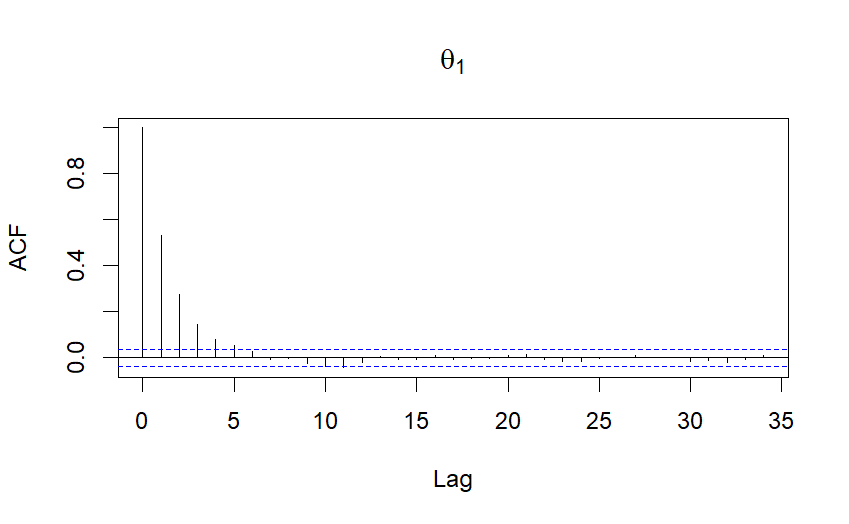}\quad
\caption{ACF plots: Lung tumor data analysis}
\label{mice_acf}
\end{figure}

\begin{figure}[h!]
\centering
\includegraphics[width=.4\textwidth]{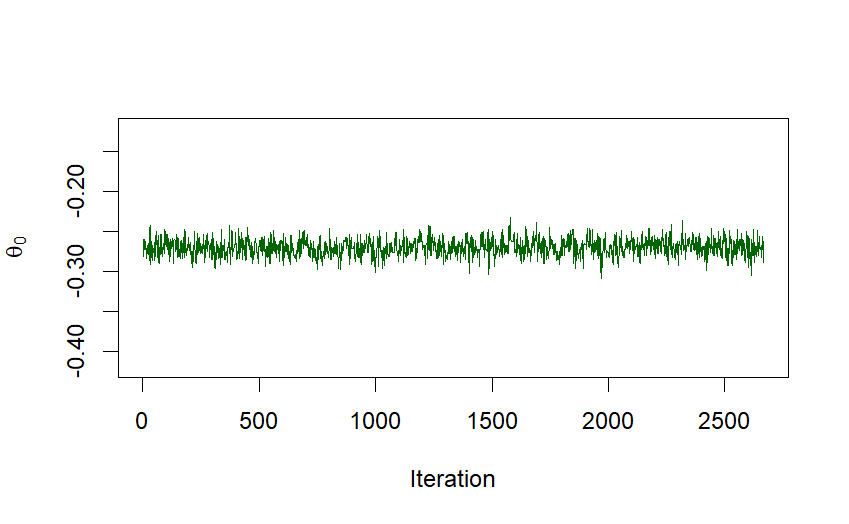}\quad
\includegraphics[width=.4\textwidth]{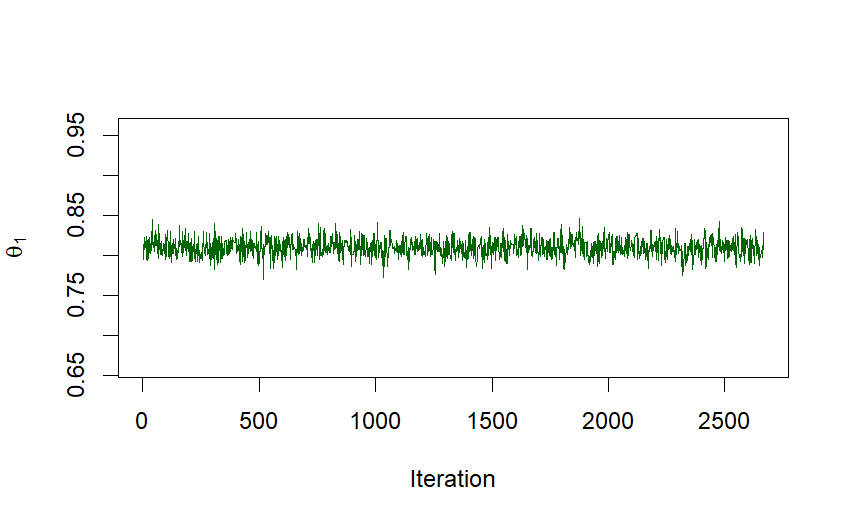}\quad
\caption{Trace plots: Lung tumor data analysis}
\label{mice_trace}
\end{figure}

\begin{figure}[h!]
\centering
\includegraphics[width=.4\textwidth]{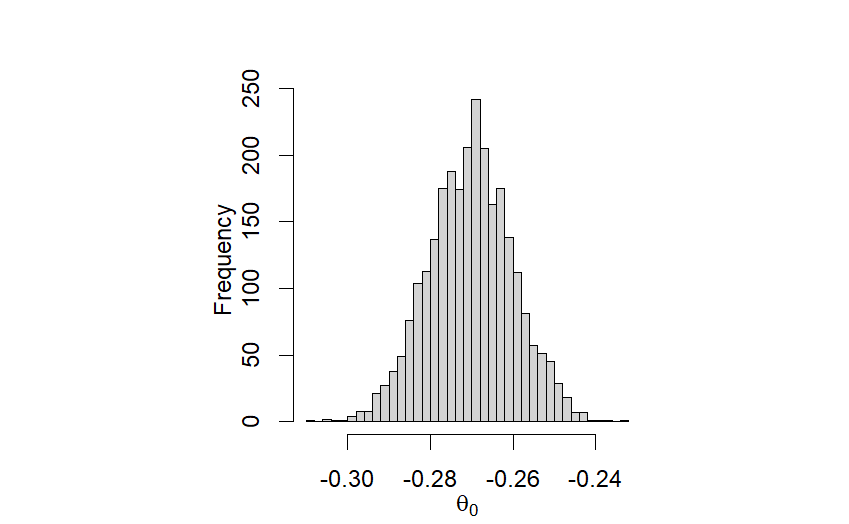}\quad
\includegraphics[width=.4\textwidth]{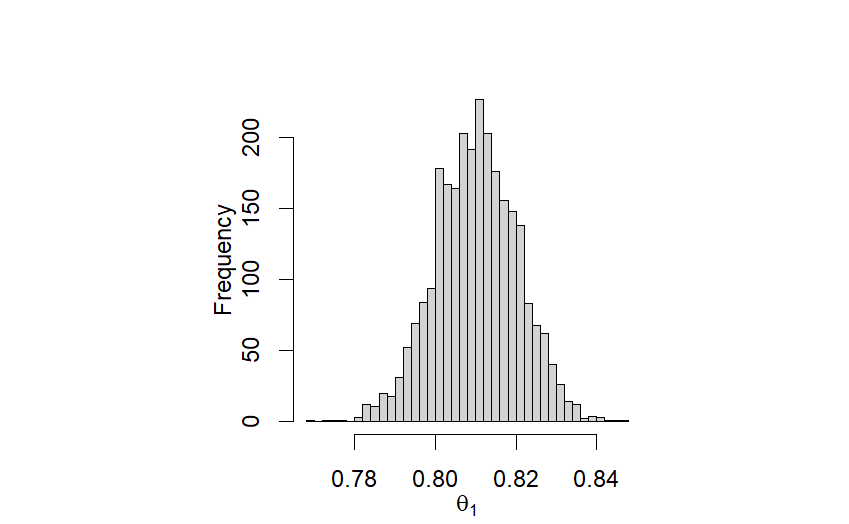}\quad
\caption{Posterior histograms: Lung tumor data analysis}
\label{mice_ph}
\end{figure}

In the analysis of lung tumor data, Markov chain diagnostics employ 50,000 MCMC samples, with a burn-in of 10,000 and retention of every 15th sample. Various graphical assessments are illustrated in Fig. \ref{mice_acf}, \ref{mice_trace}, and \ref{mice_ph}. The Gelman-Rubin diagnostic values, approaching 1, indicate that the chains have converged effectively. The ESS for $\theta_{0}$  and  $\theta_{1}$ are 399 and 500 respectively. Each iteration takes 0.0108 seconds to compute, with an acceptance rate of 0.1072.

\subsection{Breast Cancer Data Analysis }\label{A3}
\begin{figure}[h!]
\centering
\includegraphics[width=.3\textwidth]{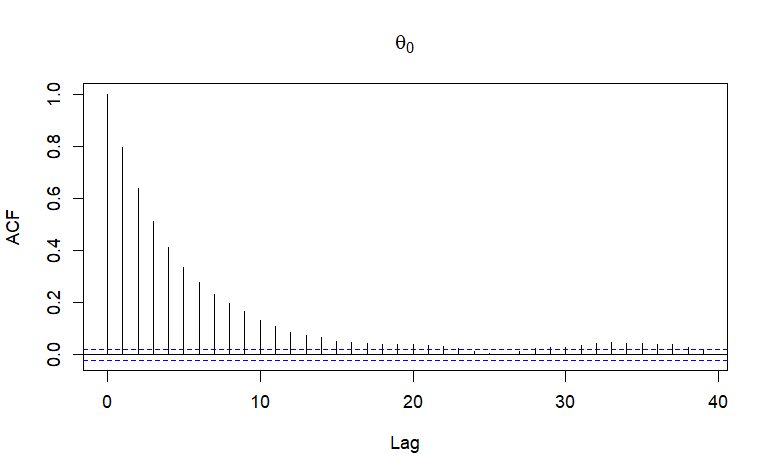}\quad
\includegraphics[width=.3\textwidth]{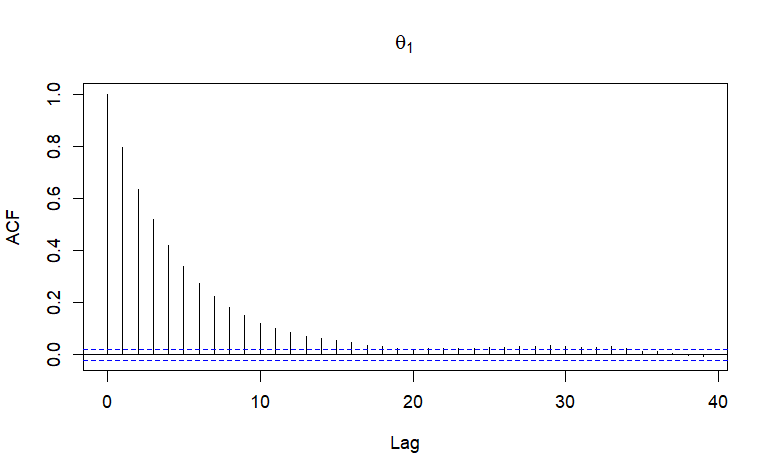}\quad
\includegraphics[width=.3\textwidth]{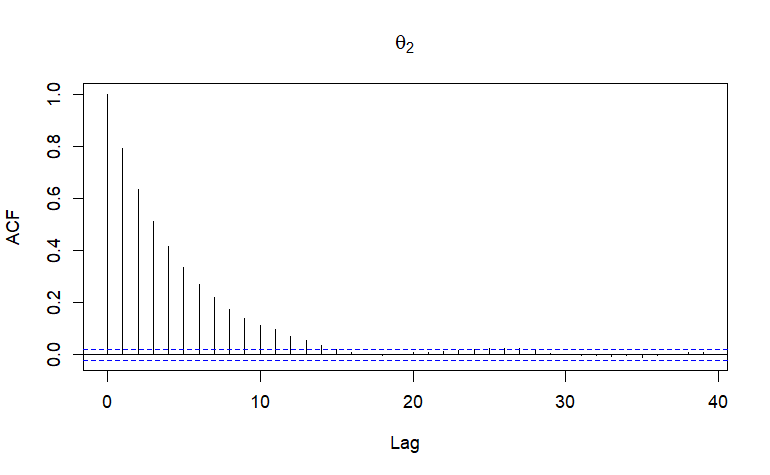}
\medskip
\includegraphics[width=.3\textwidth]{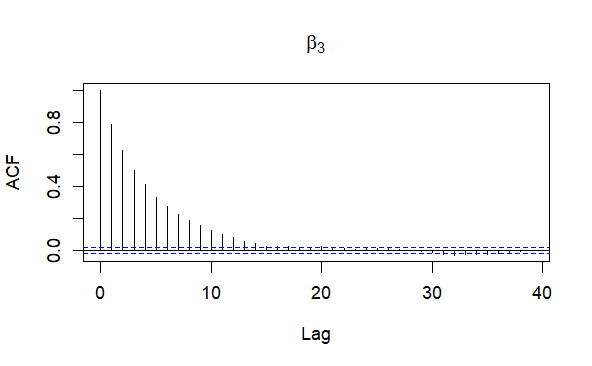}\quad
\includegraphics[width=.3\textwidth]{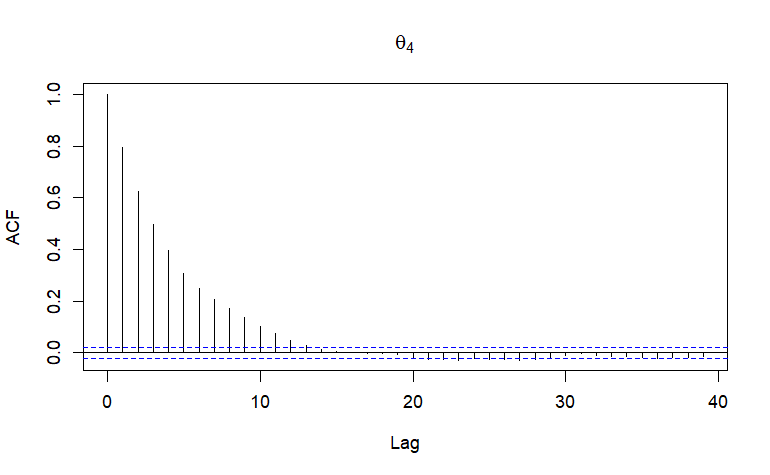}
\caption{ACF plots: Breast cancer data analysis}
\label{BC_acf}
\end{figure}

\begin{figure}[h!]
\centering
\includegraphics[width=.3\textwidth]{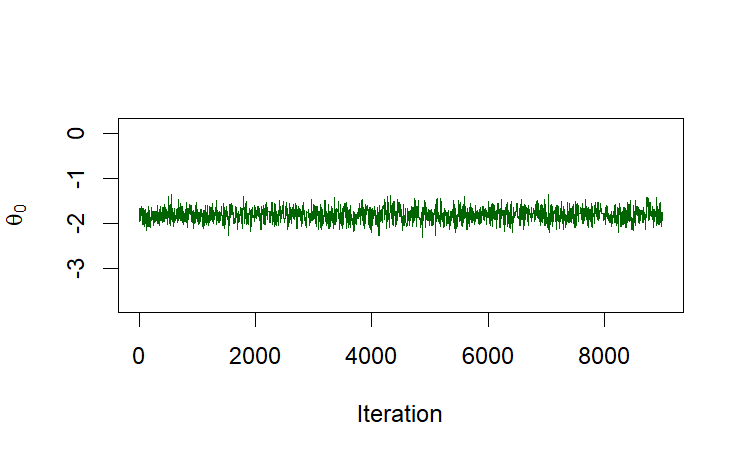}\quad
\includegraphics[width=.3\textwidth]{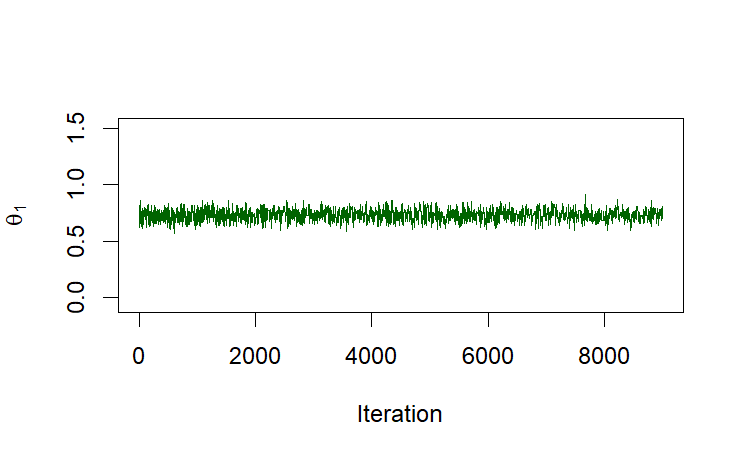}\quad
\includegraphics[width=.3\textwidth]{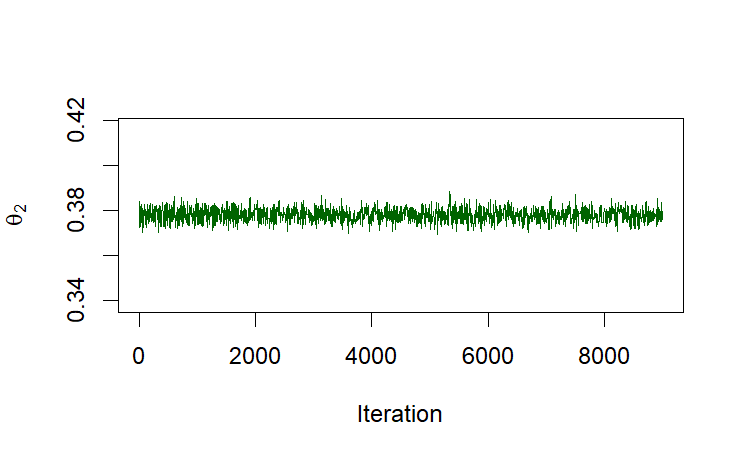}
\medskip
\includegraphics[width=.3\textwidth]{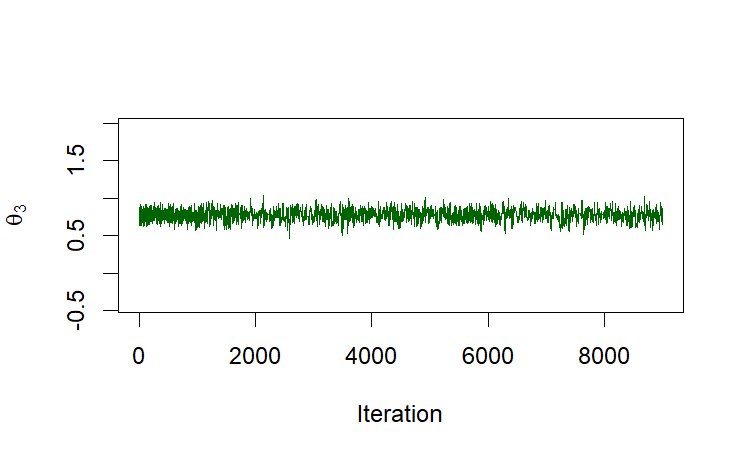}\quad
\includegraphics[width=.3\textwidth]{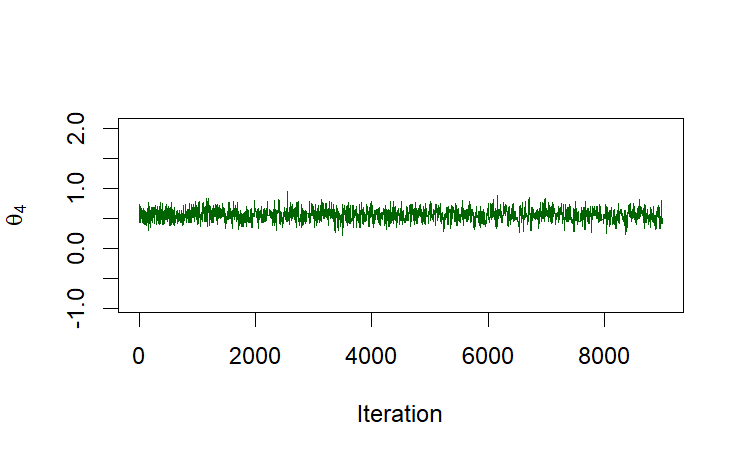}
\caption{Trace plots: Breast cancer data analysis}
\label{BC_trace}
\end{figure}

\begin{figure}[h!]
\centering
\includegraphics[width=.3\textwidth]{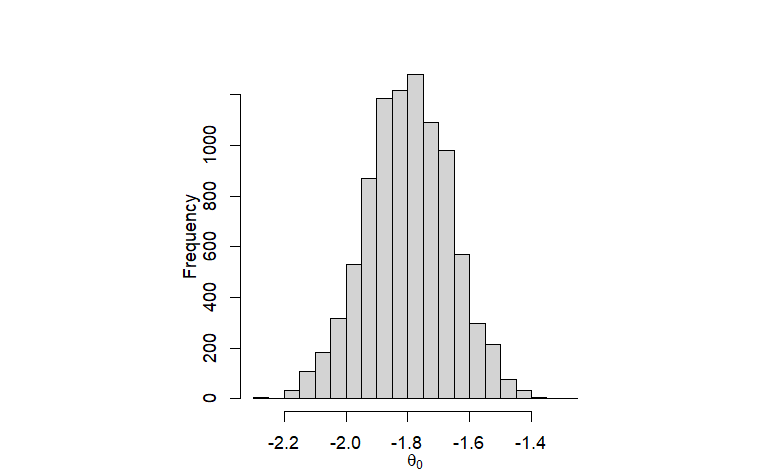}\quad
\includegraphics[width=.3\textwidth]{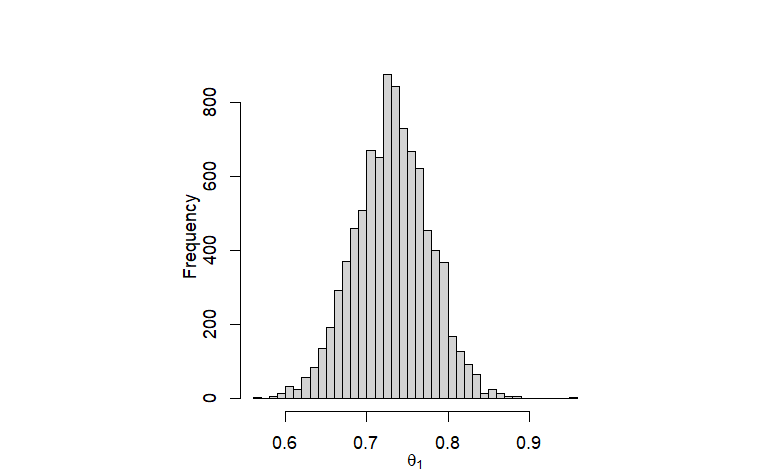}\quad
\includegraphics[width=.3\textwidth]{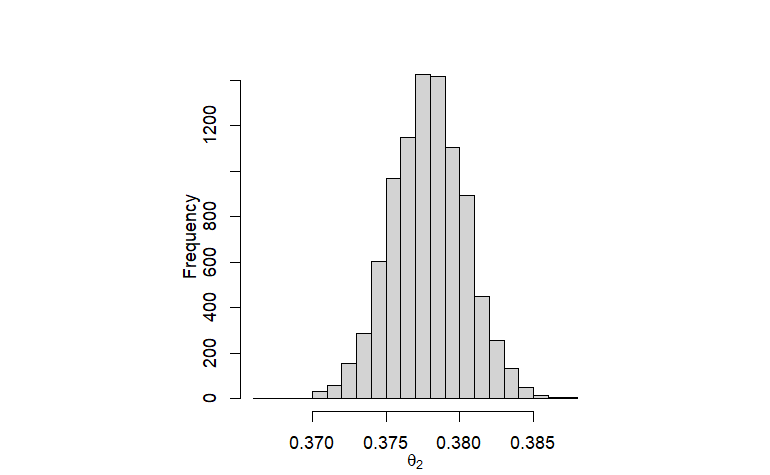}
\medskip
\includegraphics[width=.3\textwidth]{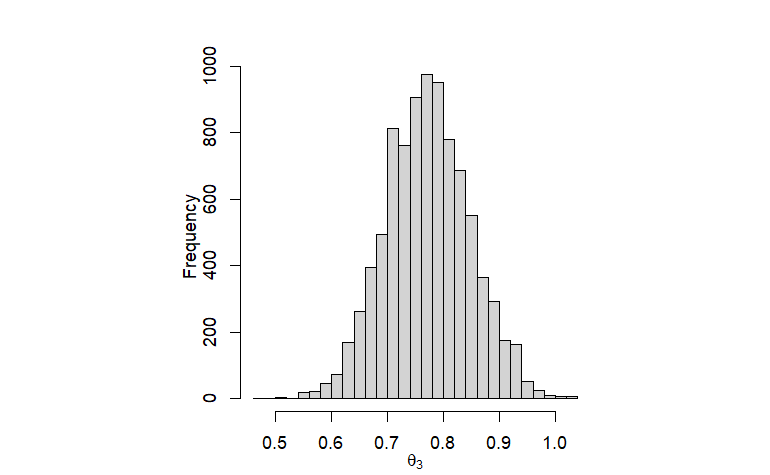}\quad
\includegraphics[width=.3\textwidth]{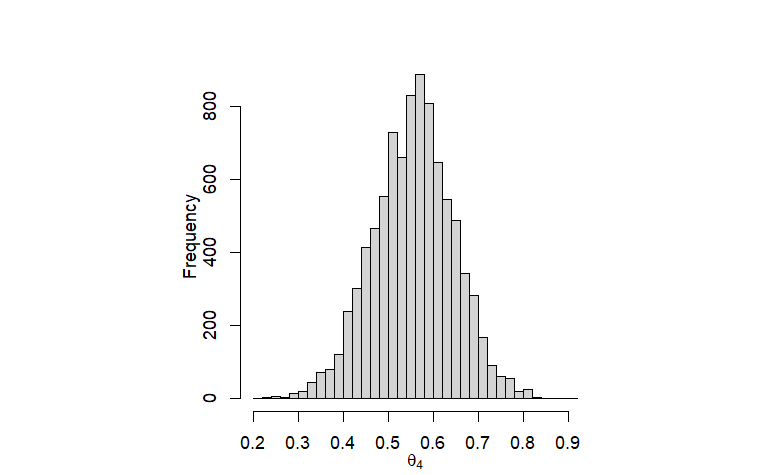}
\caption{Posterior histograms: Breast cancer data analysis}
\label{BC_ph}
\end{figure}

In the breast cancer data analysis, Markov chain diagnostics utilise 100,000 MCMC samples, with 10,000 as burn-in and retaining only the multiples of 10. Fig. \ref{BC_acf}, \ref{BC_trace}, and \ref{BC_ph} demonstrate various graphical checks. Gelman-Rubin diagnostics values close to 1 further confirm the chains' convergence. ESS for $\theta_{0}$, $\theta_{1}$, $\theta_{2}$, $\theta_{3}$,  and  $\theta_{4}$ are 708, 732, 805, 740, and 873 respectively. Computing time is 0.0152 seconds per iteration and acceptance rate is 0.0602.

\end{document}